\documentclass[a4paper,11pt]{article}
\usepackage{jcappub}
\usepackage{aas_macros}
\usepackage{lineno}
\usepackage{orcidlink}
\usepackage{bm}
\usepackage{color}
\usepackage[normalem]{ulem}

\newcommand{\lya}{Ly\ensuremath{\alpha}}
\newcommand{\poned}{$P_{\mathrm{1D}}$}
\newcommand{\pthreed}{$P_{\mathrm{3D}}$}
\newcommand{\Tr}{\mathrm{Tr}}
\newcommand{\lhf}{\mathcal{L}}

\newcommand{\partderiv}[2]{\frac{\partial #1}{\partial #2}}

\newcommand{\kms}{km~s$^{-1}$}

\newcommand{\impc}{Mpc$^{-1}$}
\newcommand{\pcm}{P$^3$M}

\newcommand{\cmnt}[1]{}

\title{Light in the dark forest -- I. 
An efficient optimal estimator for 3D Lyman-alpha forest power spectrum}

\author{Naim~G\"{o}ksel Kara\c{c}ayl{\i}\orcidlink{0000-0001-7336-8912}}
\author{and Christopher M. Hirata\orcidlink{0000-0002-2951-4932}}
\affiliation{Center for Cosmology and AstroParticle Physics, The Ohio State University,\\
191 West Woodruff Avenue, Columbus, OH 43210, USA}
\affiliation{Department of Physics, The Ohio State University,\\
191 West Woodruff Avenue, Columbus, OH 43210, USA}
\affiliation{Department of Astronomy, The Ohio State University,\\
140 W 18th Avenue, Columbus, OH 43210, USA}

\emailAdd{karacayli.1@osu.edu}
\emailAdd{hirata.10@osu.edu}

\abstract{
The highly anisotropic nature of the Lyman-alpha (\lya) forest data introduces a complex survey window function that complicates the measurement of the three-dimensional power spectrum (\pthreed). In this paper, we present the first fully optimal estimator for \pthreed, which exactly deconvolves the survey window function and marginalizes contaminated modes that distort the power spectrum. Our approach adapts optimal estimator techniques developed for the 2D cosmic microwave background data to the 3D case. To achieve computational feasibility, we employ the conjugate gradient method and implement the \pcm\ formalism to handle large-scale and small-scale operations separately and efficiently. We validate our estimator using Monte Carlo mocks and Gaussian simulations, demonstrating its accuracy and computational efficiency. We confirm that mode marginalization eliminates distortions arising from quasar continuum errors and delivers robust power spectrum estimation, though it also inflates errors at large scales. This first implementation works in the flat-sky case; we discuss the remaining steps needed to generalize it to the curved-sky case. This formalism offers a foundation for the \lya\ forest \pthreed\ measurements and a new path toward cosmological constraints from the \lya\ forest data.}

\begin{document}
\maketitle
\flushbottom
\newpage

\section{Introduction}
Neutral hydrogen is a strong absorber, capable of completely blocking light from extragalactic sources, even in trace amounts. Photons emitted blueward of the \lya\ line are redshifted as they travel through the universe and can be subjected to resonant absorption with neutral hydrogen atoms they encounter. Gunn and Peterson first demonstrated that the observed flux of 3C 9 blueward of \lya\ indicated a highly ionized intergalactic medium (IGM) \cite{schmidtQuasarRedshifts1965, gunnDensityNeutralHydrogen1965}. This finding implied that the universe underwent another major phase transition (reionization) following the earlier recombination phase transition. Over the following decades, the study of these absorption features, known as the \lya\ forest, became a powerful tool for mapping vast volumes of the universe between redshifts 2 and 5, achieving spatial resolutions as fine as kiloparsec scales.  

On large scales, the \lya\ forest technique provided measurements of the standard ruler, baryon acoustic oscillation (BAO) scale, using 3D correlation function statistics ($\xi_\mathrm{3D}$) \cite{slosarLya3DBOSS2011, slosarMeasurementBaryonAcoustic2013, buscaBaryonAcousticOscillations2013, font-riberaQuasarLymanForestCrosscorrelation2014, delubacBAOLyaDR112015, bautistajuliane.MeasurementBaryonAcoustic2017, blomqvistBaryonAcousticOscillations2019, deSainteAgatheBAOLyaDR142019, bourbouxCompletedSDSSIVExtended2020}. Recently, the measurement of the Alcock–Paczyński effect enabled a ``full-shape" analysis and measurement of $\Omega_m$ from the \lya\ forest data alone \cite{cuceuConstraintsCosmicExpansion2023}. On smaller scales, statistics of interest have been the 1D power spectrum (\poned), which has access to small-scale physics such as the thermal state of the IGM \cite{boeraRevealingThermal2019, waltherNewConstraintsIGM2019, villasenorThermalHistory2022}, the suppression due to the mass of neutrinos \cite{croftNeutrinoMassLyaForest1999, palanqueDelabrouilleNeutrinoMass2015, yecheNeutrinoMassesXQ2017} and dark matter particles \cite{narayananWDMLyaForest2000, seljakSterileNeutrinosDM2006, boyarskyLyaWDM2009, wangLyaDecayingDM2013, vielWarmDarkMatter2013, irsicFuzzyDMfromLya2017, irsicConstraintsWDM2017, villasenorWarmDarkMatter2023}.

Current and next-generation surveys will significantly increase the amount of \lya\ forest data. The Dark Energy Spectroscopic Instrument \cite[DESI,][]{leviDESIExperimentWhitepaper2013} became operational in 2021 with a mission to collect 40 million spectra from galaxies and quasars in five years. The five-year DESI data will observe the \lya\ forest from approximately 700,000 quasar spectra at $z \gtrsim 2.1$ \cite{desicollaborationDESIExperimentPart2016, desicollaborationDESIExperimentPart2016b}, which will be four times larger than the preceding experiment, the Extended Baryon Oscillation Spectroscopic Survey \citep[eBOSS,][]{dawsonSDSSeBOSSEarlyData2016}.

This great increase rekindled efforts to measure the 3D power spectrum of the \lya\ forest (\pthreed) \cite{fontriberaEstimate3DPowerLya2018, abdulkarimMeasurementofSmallScaleP3D2024, belsunce3dLymanAlphaPowerSpectru2024, horowitzMapleLya3Destimator2025}. In principle, the large-scale 3D power spectrum contains a wealth of information about ionizing background fluctuations \cite{pontzenScaleDependentBiasIGM2014, gontchoAGontchoEffectOfIonizingBackgroundLya2014, meiksinTimeDependentFluctuations2019, longProbingLargeScaleIonizingBackgroundLya2023}, hydrogen \cite{monteroCamachoImpactOfInhomogeneousReionization2019, monteroCamachoLyaPowerSpectrumWindowReionization2020} and helium \cite{mcquinnSignaturesOfLargeScaleTemperature2011} reionization, pre-reionization thermal history \cite{monteroCamachoLongLastingEffectXrayIGM2024}, and primordial non-Gaussianity \cite{dalalImprintsOfPrimordialNongaussianities2008, seljakBiasRsdPrimordialNongaussianity2012}. Even though the theoretical \pthreed$(\bm k)$ and $\xi_\mathrm{3D}(\bm r)$ are Fourier transform pairs and equivalent in information content, the same is not true for the estimated $\hat P_\mathrm{3D}(\bm k)$ and $\hat \xi_\mathrm{3D}(\bm r)$ from noisy observations \cite{feldmanPowerSpectrumAnalysis1994}. For example, the errors of $\hat P_\mathrm{3D}(\bm k)$ are significantly less correlated between $\bm k$ modes compared to the errors of $\hat \xi_\mathrm{3D}(\bm r)$ between $\bm r$ bins. The survey window function mixes different  $\bm k$ modes in $\hat P_\mathrm{3D}(\bm k)$, whereas $\hat \xi_\mathrm{3D}(\bm r)$ is robust against this effect. In fact, the strongly anisotropic nature of \lya\ forest data --- sparse sampling in the transverse direction and dense (but noisy) sampling in the line-of-sight direction --- introduces a highly complex, challenging survey window function to \pthreed\ estimation methods. The workaround to this problem has been to (1) propose an alternative hybrid statistic called $P_\times(k_\|, \Delta \theta)$ that is in Fourier space in the line-of-sight direction and configuration space in the transverse direction \cite{fontriberaEstimate3DPowerLya2018, abdulkarimMeasurementofSmallScaleP3D2024} or (2) sub-optimally estimate \pthreed\ using a weighted pair-count estimator \cite{philcoxComputingSmallScaleGalaxyPower2020, belsunce3dLymanAlphaPowerSpectru2024}.


In other words, the field has lacked an efficient, optimal estimator technique for \pthreed\ until now. The optimal estimator formalism has been developed for cosmic microwave background (CMB) maps and fruitfully applied to those data sets \cite{hamiltonOptimalMeasurementPower1997, ohEfficientPowerSpectrumCmb1999, hirataCrossCorrelationCmbLss2004}. \poned\ of the \lya\ forest has been optimally estimated by refs.~\cite{mcdonaldLyUpalphaForest2006, karacayliOptimal1DLy2020} thanks to its reduced dimensionality. These optimal estimators minimize error bars, exactly deconvolve the survey window function, and marginalize contaminated modes from the data, making them highly desirable. Specifically, in the current \lya\ forest analyses, the errors in the quasar continuum distort the two-point statistics and significantly complicate the cosmological inferences. Our optimal estimator not only yields the smallest error bars but also eliminates these distortions from the measurement.

At its core, the optimal estimator is a weighted average, where the weights are given by the inverse covariance matrix of the entire data vector. This makes the core computational operation $\sim \mathbf{C}^{-1} \bm x$. However, for data vectors of size $\mathcal{O}(10^{7-9})$, directly computing this inverse is infeasible. Instead of explicitly inverting the covariance matrix, we solve the linear system $\mathbf{C} \bm y = \bm x$ for $\bm y$ using the conjugate gradient method, as is commonly done in CMB analyses \cite{ohEfficientPowerSpectrumCmb1999}. Moreover, the operation $\mathbf{C} \bm y$ becomes computationally intractable if all matrix-vector multiplications are explicitly computed for every forest pair. We make this operation feasible with the \pcm\ formalism, originally developed for numerical gravity solvers and later applied to weak lensing surveys in optimal estimator formalism \cite{padmanabhanMiningWeakLensing2003}. In this approach, the large-scale matrix-vector multiplication ($\mathbf{C}_L \bm y$) is performed using a mesh-based method, capitalizing on fast Fourier-space convolution (particle-mesh, PM). Meanwhile, the small-scale matrix-vector multiplication ($\mathbf{C}_S \bm y$) is computed exactly through direct summation between neighboring forests (particle-particle, PP). 

In this work, we describe our proposed optimal estimator algorithm and its implementation in a Cartesian grid while focusing on overcoming \lya\ specific challenges. We address mode marginalization, redshift evolution along the radial direction, and generating Monte Carlo simulations with proper correlations between lines of sight. The treatment of spherical geometry is left for future work. Our approach provides a robust framework for estimating the \lya\ forest \pthreed\ with good accuracy and high efficiency, and paves the way for precise cosmological measurements.

This algorithm can optimally extract information from both large, quasi-linear scales and small, non-linear scales. In this work, however, we focus on the large scales, where the estimator proves most impactful for sparsely sampled, large-scale structure surveys such as DESI. Our method can recover full-shape information in the power spectrum, revealing Alcock–Paczyński distortions, ionizing background fluctuations, and signatures of primordial non-Gaussianity.  Conversely, a densely sampled survey like the Ly$\alpha$ Tomography IMACS Survey (LATIS, \cite{newmanLatisLyaTomography2020}) is better suited for measuring small-scale \pthreed, though the benefits of the optimal estimator formalism may be limited unless mode marginalization becomes necessary.

This paper is organized as follows. Section~\ref{sec:setup} presents the setup of our problem, including the geometry and the input \pthreed\ model to the optimal estimator, where a fitting function is sufficient since the estimator only needs to approximate the underlying signal. Section~\ref{sec:optimal_estimation}, the central section of this paper, provides a detailed description of our optimal estimator. Validation tests using Monte Carlo mocks and Gaussian simulations are discussed in section~\ref{sec:validation}. The advantages and limitations of our method, and directions for future work, are examined in section~\ref{sec:discussion}.

\section{Setup}
\label{sec:setup}

We now set up the power spectrum estimation problem addressed in this paper. This includes a model for the fiducial \lya\ power spectrum (section~\ref{sec:model}) and the simplified geometry within the flat-sky approximation (section~\ref{sec:geometry}).

\subsection{P3D model}
\label{sec:model}
In linear theory, the \lya\ forest power spectrum can be modeled with a linear bias term and a Kaiser term for the redshift space distortions. A major complication of modeling \pthreed\ is the non-linear corrections $D_\mathrm{NL}(k, \mu)$ to this linear model. We further need to introduce an additional suppression to prevent small-scale modes from aliasing.
\begin{equation}
    P_\mathrm{3D}(k, \mu) = b_F^2 (1 + \beta_F \mu^2)^2 P_L(k) D_\mathrm{NL}(k, \mu) e^{-k^2/k_A^2},
\end{equation}
where $P_L(k)$ is the linear matter power spectrum at redshift $z$, and the additional exponential suppression at a characteristic scale $k_A=0.4$~\impc\ is for numerical stability. To be consistent across our comparisons, we include this numerical term in our model. Empirical fitting functions of $D_\mathrm{NL}(k, \mu)$ aim to capture three major deviations from the linear theory: non-linear enhancement, pressure smoothing, and line-of-sight broadening. Non-linear collapse enhances the power. The thermal broadening of line profiles and non-linear peculiar velocities smooth the power spectrum only in the line-of-sight direction. Lastly, the gas pressure isotropically suppresses the power at a characteristic scale known as the filtering scale or Jeans scale \cite{gnedinProbingUniverseLya1998}. As we will see below, this effect is negligible compared to the smoothing required for numerical stability and to prevent aliasing.

We use the fitting function for $D_\mathrm{NL}(k, \mu)$ presented in ref.~\cite{arinyoNonLinearPowerLya2015} based on simulations, which is inspired by and improved upon the work in ref.~\cite{mcdonaldPredictingLyaPower2003} by reducing the number of free parameters to 5.
\begin{equation}
    \ln D_\mathrm{NL}(k, \mu) = q_1 \Delta^2(k) \left[1 - (k / k_\nu)^{a_\nu} \mu^{b_\nu} \right] - ( k / k_p)^2,
\end{equation}
where $\Delta^2(k) = k^3 P_L(k) / 2\pi^2$. The non-linear enhancement is quantified by the parameter $q_1$. The pressure smoothing is captured by the parameter $k_p$. The anisotropic line-of-sight broadening is described with additional parameters that multiply the non-linear enhancement term.

Ref.~\cite{chabanierAccel2Simulations2024} has simulated the \lya\ forest using state-of-the-art techniques and provides the most recent best-fitting parameters. To quantify the redshift evolution of the bias, $b_F(z)$, we fit a power law $b_F(z) = b_F^\mathrm{pivot} (\frac{1 + z}{1 + z_\mathrm{pivot}})^{\alpha_F}$ to the values listed in Table A3 of ref.~\cite{chabanierAccel2Simulations2024}, with $z_\mathrm{pivot} = 2.4$. The remaining parameters are linearly interpolated to $z_\mathrm{pivot}$. The values used in the optimal estimator are: $b_F^\mathrm{pivot}=-0.1196$, $\alpha_F=3.38$, $\beta_F=1.66$, $q_1=0.796$, $k_\nu=0.392$~Mpc$^{-1}$, $a_\nu=0.383$, $b_\nu=1.65$. The reported value of $k_p=16.8$~\impc\ can be ignored since $1/k_A \gg 1/k_p$.


The linear power spectrum is computed using the Boltzmann solver \texttt{CAMB} \cite{lewisCAMB2000}, assuming a flat $\Lambda$CDM universe with Planck 2018 cosmology \cite{collaborationPlanck2018Results2020} at the pivot redshift.

\subsubsection{High-column density systems}
High-column density (HCD) systems trace highly overdense regions of the underlying matter field and follow a different bias relation, while their saturated absorption profiles strongly affect forest fluctuations along the line of sight. Since it is nearly impossible to completely identify, mask, and correct for these saturated absorption regions, HCDs inevitably modify the correlation function and power spectrum. 

This modification follows a simple form at linear order \cite{font-riberaEffectHighColumn2012, rogersCorrelations3dHighColumnDensity2018, bourbouxCompletedSDSSIVExtended2020}. Using the definitions $\tilde b_F = b_F (1 + \beta_F \mu^2)$ and $\tilde b_\mathrm{HCD} = b_\mathrm{HCD} (1 + \beta_\mathrm{HCD} \mu^2) F_\mathrm{HCD}(k_\|)$, the total 3D power spectrum is expressed as
\begin{equation}
    P_\mathrm{3D}(k, \mu) = P_L(k) \left(\tilde b_F^2 D_\mathrm{NL}(k, \mu) + A_\mathrm{halo}(k) (2\tilde b_F \tilde b_\mathrm{HCD} + \tilde b_\mathrm{HCD}^2) \right),
\end{equation}
where $b_\mathrm{HCD}$ and $\beta_\mathrm{HCD}$ are the bias parameters for the absorption caused HCDs. We initially introduced the term $A_\mathrm{halo}$ to capture the decorrelation between the \lya\ forest and haloes at $1.5~$\impc\ \cite{givansNonlinearLyaCross2022}.\footnote{This decorrelation should not be included in the autocorrelation term $\tilde b_\mathrm{HCD}^2$.} However, aliasing occurs at larger scales than this decorrelation. So $A_\mathrm{halo}$ is repurposed to suppress small-scale power to prevent aliasing. \footnote{In hindsight, this aliasing term could have been combined with the one above to have a single exponential suppression.}
Lastly, we use the approximation $F_\mathrm{HCD}(k_\|) = \exp(-L_\mathrm{HCD} k_\|)$, which is the Fourier transform of the Lorentzian broadening in the Voight profile, and $L_\mathrm{HCD}$ is the typical length of the HCD absorption \cite{bourbouxCompletedSDSSIVExtended2020}.

We cut off contributions from HCDs at $k_\mathrm{max} = 0.4~$\impc, to mitigate aliasing and inaccurate modelling at small scales. To smoothly suppress these small-scale contributions, we apply an apodizing function identical to that in section~\ref{subsec:computing_sz}:
\begin{align}
    A_\mathrm{halo}(k) &= \begin{cases} 
          1 & k < k_\mathrm{max} / 2\\
          \cos^2\left(\frac{\pi}{2} (\frac{k}{k_\mathrm{max} / 2} - 1 )\right) & k_\mathrm{max} / 2 \leq k \leq k_\mathrm{max} \\
          0 & k_\mathrm{max} < k
       \end{cases},
\end{align}
Our HCD model adopts best-fit parameters from ref.~\cite{bourbouxCompletedSDSSIVExtended2020}: $b_\mathrm{HCD} = -0.05$, $\beta_\mathrm{HCD} = 0.7$, and $L_\mathrm{HCD} = 14.8$~Mpc.

\subsubsection{Metals}
The effect of non-Lya absorbers in the forest is small but non-negligible. Ionized silicon has transition wavelengths close to the \lya\ wavelength and adds oscillations (peaks) to the power spectrum (correlation function) through redshift confusion. Let us illustrate how these systems can be included in our \pthreed\ estimation for the strongest line, Si~\textsc{iii} (1206.52~\AA).

First, the redshift confusion places Si~\textsc{iii} systems at an offset $\Delta r = \chi(z_\mathrm{SiIII}) -  \chi(z_\mathrm{\lya})$, where $(1 + z_\mathrm{SiIII}) \lambda_\mathrm{SiIII} = (1 + z_\mathrm{\lya}) \lambda_\mathrm{\lya}$. With cosmological parameters above, $\Delta r\approx 30~$Mpc at $z_\mathrm{\lya} = 2.4$. In general, $\Delta r$ changes with redshift, but this change is negligible since $\Delta r(z=2.7) - \Delta r(z=2.1) \approx 2~$Mpc. The peak in the correlation function at this separation is smeared due to this effect, redshift space distortions, and clustering. As these ions reside in the densest regions of the universe, we again refer to the \lya\ forest-halo cross-correlation study of ref.~\cite{givansNonlinearLyaCross2022}. We model the redshift space distortions with a Lorentzian decay $D_\mathrm{FoG}=1 / 1 + (k_\| \sigma_v)^2$ and suppress the contribution of metals with the same $A_\mathrm{halo}(k)$ function. Similarly, defining $\tilde b_\mathrm{SiIII} = b_\mathrm{SiIII} (1 + \beta_\mathrm{SiIII} \mu^2)$, the Si~\textsc{iii} contribution to the total \pthreed\ is
\begin{equation}
    P_L(k) A_\mathrm{halo}(k)\left(2\tilde b_F \tilde b_\mathrm{SiIII} \cos(k_\| \Delta r) \sqrt{D_\mathrm{NL} D_\mathrm{FoG} }+ \tilde b_\mathrm{SiIII}^2 D_\mathrm{FoG} \right),
\end{equation}
where we use $\beta_\mathrm{SiIII}=0.5$ and $\sigma_v=5~$Mpc \cite{bourbouxCompletedSDSSIVExtended2020, desiKp6BaoLya2024}.\footnote{This $\sigma_v$ is approximately the best-fitting Lorentzian parameter value for the quasar fingers of god effect in ref.~\cite{desiKp6BaoLya2024} and may be non-negligibly influenced by quasar redshift errors. We only need to approximate this smoothing in our model, so it is sufficient for our purposes.}

In the correlation function analysis of DESI, all three Si~\textsc{ii} lines and a Si~\textsc{iii} line at 1206.52~\AA\ are modeled, including their cross-correlations with each other \cite{desiKp6BaoLya2024}. We include all these in our metal contamination model with the following parameters: $b_\mathrm{SiIII(1207)} = 9.8\times 10^{-3}$, $b_\mathrm{SiII(1190)} = 4.5\times 10^{-3}$, $b_\mathrm{SiII(1193)} = 3.05\times 10^{-3}$, and $b_\mathrm{SiII(1260)} = 4.0\times 10^{-3}$.



\subsection{Geometry}
\label{sec:geometry}
For this first implementation, we work in Cartesian geometry with a flat sky. This is the first step in understanding the fundamental linear algebra issues before proceeding to the spherical geometry of a real survey.

Unrolling spherical shells into Cartesian space inevitably distorts the geometry. To prevent sightlines from diverging, we fix $(x, y)$ coordinates at an effective comoving distance $\chi_\mathrm{eff}$ such that:
\begin{equation}
    x = \chi_\mathrm{eff} (\phi + \phi_0), \quad  y = \chi_\mathrm{eff} (\theta - \theta_0), \quad z = \chi(z_r) - \chi_0 
\end{equation}
where $\theta$ is declination, $\phi$ is right ascension of the quasar, and $\chi(z_r)$ is the comoving distance to redshift $z_r$. $\chi_0$ is the initial distance in the radial direction where the mesh begins and is chosen to be large enough to minimize the periodicity effects from FFT. $\theta_0$ is chosen as the median declination to further minimize distortions due to projection. 

\section{Optimal estimation of P3D}
\label{sec:optimal_estimation}
Our main task is to optimally estimate the 3D power spectrum of the \lya\ forest on large, weakly non-linear scales from noisy and spatially sparse transmitted flux fluctuations. The quasar spectra with the \lya\ forest data form unevenly sampled, infinitesimal lines in space. The transmitted flux fluctuations contain nonuniform noise and are biased due to the quasar continuum determination procedure that fits the underlying large-scale fluctuations. The data vector can be expressed as $\bm{d} = \bm\delta_F + \bm n + \bm u$, where $\bm\delta_F$ are the transmitted flux fluctuations convolved by the spectrograph window function in the line of sight, $\bm n$ is noise, and we denote the continuum error templates by vector $\bm u$.

The quadratic (optimal) estimator formalism has been a highly successful technique in power spectrum measurements from the cosmic microwave background radiation, galaxy surveys, and weak lensing \cite{hamiltonOptimalMeasurementPower1997, tegmarkKarhunenLoeveEigenvalueProblems1997, tegmarkMeasuringGalaxyPower1998, seljakWeakLensingReconstruction1998, ohEfficientPowerSpectrumCmb1999, padmanabhanMiningWeakLensing2003}. We use this formalism to build a \pthreed\ estimator that maximizes the likelihood function for the data $\bm d$ under the Gaussianity assumption. We use $\lhf$ to denote the negative logarithm of the probability, 
    $2\mathcal{L}(P_\mathrm{3D}(k_\bot, k_\|) | \bm d) =\ln\det\mathbf{C} + \bm{d}^\mathrm{T}\mathbf{C}^{-1} \bm{d}$,
where $\mathbf{C} = \langle \bm d \bm d^\mathrm{T} \rangle = \mathbf{S} + \mathbf{N}$ is the full covariance matrix. The most likely \pthreed\ can be found by maximizing this likelihood function: $\partial \lhf / \partial P_k=0$ (we use subscript $k$ to denote one $(k_\bot, k_\|)$ bin for clarity from here on), which can be iteratively solved using the Newton-Raphson method. However, with a good initial estimate of the covariance matrix, a single iteration is sufficient \cite{karacayliOptimal1DLy2020}. The maximum likelihood estimator is then
\begin{equation}
    \label{eq:p3d_estimator} \hat P^\mathrm{3D}_{k} = P^\mathrm{3D, fid}_{k} + \sum_{k'} \frac{1}{2} F^{-1}_{kk'}\left[\bm{d}^\mathrm{T} \mathbf{C}^{-1}\mathbf{C}_{,k'} \mathbf{C}^{-1} \bm{d} -  \mathrm{Tr}(\mathbf{C}^{-1}\mathbf{C}_{,k'}) \right],
\end{equation}
where $\mathbf{C}_{,k}\equiv\partial\mathbf{C}/\partial P^\mathrm{3D}_{k}$ and $\mathrm{Tr}(\mathbf{C}^{-1}\mathbf{C}_{,k})$ is the bias term, and the estimated Fisher matrix is
\begin{equation}
    \label{eq:fisher_matrix}F_{kk'} \equiv \left\langle \partderiv{^2 \lhf}{P^\mathrm{3D}_{k} \partial P^\mathrm{3D}_{k'}} \right\rangle = \frac{1}{2} \Tr(\mathbf{C}^{-1}\mathbf{C}_{, k} \mathbf{C}^{-1} \mathbf{C}_{, k'} ).
\end{equation}
Thus, computing $\mathbf{C}$, its inverse, and its derivatives with respect to \pthreed\ is the focal point of this method.

\subsection{Conjugate gradient solution of \texorpdfstring{$\mathbf{C}^{-1} \bm x$}{C-1x}}
The data vector $\bm d$ has $3.4\times 10^7$ pixels for coarse-grained eBOSS data and is expected to grow to $6\times 10^8$ for full-resolution DESI data.\footnote{This large increase in number of pixels cannot be explained by improved resolution of approximately $20-30\%$ in DESI and likely comes from improvements in the rest of the pipeline.} The covariance matrix $\mathbf{C}$ cannot be stored in memory or inverted trivially.
We instead solve $\bm y = \mathbf{C}^{-1} \bm x$ iteratively using the conjugate gradient method.

We first put the equation $(\mathbf{S} + \mathbf{N}) \bm y = \bm x$ into a more stable form while retaining the symmetric, positive-definite structure:
\begin{align}
    (\mathbf{I} + \mathbf{N}^{-1/2} \mathbf{S} \mathbf{N}^{-1/2}) \mathbf{N}^{1/2} \bm y &= \mathbf{N}^{-1/2} \bm x \\
    \label{eq:cgd_matrix_zm}(\mathbf{I} + \mathbf{N}^{-1/2} \mathbf{S} \mathbf{N}^{-1/2}) \bm z &= \bm m.
\end{align}
where we start with an initial guess $\bm z^{(0)}$, and iterate until $\bm z$ converges. Then, the solution is $\bm y = \mathbf{N}^{-1/2} \bm z$ at convergence. We underscore that we only use the numerically stable $\mathbf{N}^{-1/2}$, which is zero for masked pixels.


The preconditioned conjugate gradient method requires at least three additional vectors to be stored: the residual vector $\bm r$, the search direction $\bm p$, and, of course, the output vector. After the initial guess $\bm z^{(0)}$, the residual vector is set to $\bm r^{(0)} = \bm m - \mathbf{A}\bm z^{(0)}$, where $\mathbf{A}$ is the matrix in parantheses in eq.~\eqref{eq:cgd_matrix_zm}, and the search direction is set to $\bm p^{(0)} = \mathbf{E} \bm r^{(0)}$, where $\mathbf{E}$ is the preconditioner. For iteration $(i + 1)$, we first calculate $r^{(i)}_2 = \bm r^{(i) \mathrm{T}} \mathbf{E} \bm r^{(i)}$, $\bm{\tilde{p}}^{(i)} = \mathbf{A}\bm p^{(i)}$ and $p^{(i)}_2 = \bm {\tilde{p}}^{(i) \mathrm{T}} \bm p^{(i)}$. Then, we update the solution $\bm z^{(i + 1)} = \bm z^{(i)} + \alpha^{(i)} \bm p^{(i)}$ and the residual vector $\bm r^{(i + 1)} = \bm r^{(i)} - \alpha^{(i)} \bm{\tilde{p}}^{(i)}$, where $\alpha^{(i)} = r^{(i)}_2 / p^{(i)}_2$. If convergence has not been achieved, we update the search direction $\bm p^{(i + 1)} = \mathbf{E} \bm r^{(i + 1)} + \beta^{(i)} \bm p^{(i)}$, where $\beta^{(i)} = r^{(i + 1)}_2 / r^{(i)}_2$. We end the iterations if $\| \bm r^{(i + 1)}\|_2 < \epsilon \cdot \| \bm r^{(0)}\|_2$, where we pick $\epsilon=10^{-6}$--$10^{-5}$. The algorithm typically achieves convergence between 30 and 50 iterations.

We use the block diagonal approximation as the preconditioner, $\mathbf{E}=\mathbf{A}^{-1}_\mathrm{BD}$, which is also used for the initial guess $\bm z^{(0)}=\mathbf{A}^{-1}_\mathrm{BD} \bm m$. Better preconditioners can be constructed by formulating the noise matrix in Fourier space \cite{ohEfficientPowerSpectrumCmb1999, hirataCrossCorrelationCmbLss2004}.

\subsection{Computing \texorpdfstring{$\mathbf{S} \bm z$}{Sz}\label{subsec:computing_sz}}
We use a \pcm\ algorithm to efficiently construct and perform required matrix operations following a similar technique applied to weak lensing surveys \cite{padmanabhanMiningWeakLensing2003}. We decompose the power spectrum into large- and small-scale components by an adjustable large-scale radius parameter $R_L$:
\begin{equation}
    \label{eq:P_LS}P_L = P_\mathrm{3D}(k, k_\|) e^{-k^2 R_L^2}\quad~~~{\rm and}~~~ \quad P_S = P_\mathrm{3D}(k, k_\|) (1 - e^{-k^2 R_L^2}).
\end{equation}
We perform the large-scale signal matrix $\mathbf{S}_L$ multiplication as a convolution on the mesh in Fourier space using FFTs and perform the small-scale signal matrix $\mathbf{S}_S$ multiplications only for the neighboring lines of sight for each quasar within $f R_L$, where $f$ is the neighboring radius factor that is also adjustable. In this work, we adopt values of $R_L=6$ and $8$~Mpc and $f=8$ and $12$.

\subsubsection{Large-scale component}
The large-scale part ${\bf S}_L$ uses harmonic space, and as such, it is the step that depends most sensitively on the choice of the Cartesian/flat sky (with Fourier modes) versus the curved sky (with spherical harmonics). We first test our method with Cartesian grids, which are more straightforward to implement and interpret, and leave the spherical sky treatment to future work.

We reverse interpolate vectors using the clouds-in-cell (CIC) assignment function onto a Cartesian grid that is an equirectangular projection. We then perform FFT, multiplication of each Fourier mode by $P_L({\bf k})$, FFT back to real space, and do trilinear interpolation to the sightlines. The operation ${\bf S}_L$ computed this way is semi-positive definite since $P_L({\bf k})\ge 0$, and exactly symmetric since the CIC operation from sightline pixels to Cartesian grid points is the transpose of trilinear interpolation.

\subsubsection{Small-scale component}
The small-scale correlation function is given by
\begin{equation}
    \label{eq:xi_S_dht}\xi_S(r_\bot, r_\|) = A(r) \int_0^\infty \frac{k_\bot \mathrm{d}k_\bot}{2 \pi} J_0(k_\bot r_\bot) \int_0^\infty \frac{\mathrm{d}k_\|}{\pi} \cos(k_\| r_\|) P_S(k_\bot, k_\|),
\end{equation}
where an apodizing function $A(r)$ is applied to reduce ringing in the Fourier space of a sharp transition to zero at $r=fR_L$. This spherical top-hat induced ringing can lead to negative eigenvalues in the covariance matrix. A smooth transition function suppresses this ringing, although it does not eliminate it. The apodizing function used for this purpose is given by
\begin{align}
    A(r) &= \begin{cases} 
          1 & r < r_1 \\
          \cos^2\left(\frac{\pi}{2} \frac{r - r_1}{f R_L - r_1} \right) & r_1 \leq r \leq f R_L \\
          0 & f R_L < r 
       \end{cases},
\end{align}
which smoothly transitions from one to zero starting at the breakpoint $r_1 = (3/4) f R_L$.

We compute the integrations using the \texttt{FFTLog} algorithm \cite{talmanNumericalFourierBessel1978, hamiltonUncorrelatedModesNonlinear2000}, noting that $\cos(x) = J_{-1/2}(x)$ $\sqrt{\pi x/2}$.
To reduce \texttt{FFTLog} ringing, the transformations are performed over a wide $k$ range of $10^{-6}-10^6~$\impc\ with 2048 points,\footnote{$10^6~$Mpc $ = 1~$Tpc.} and only the range $r \in [5\times 10^{-4}-10^{3}]$~Mpc is kept.\footnote{Note that there are two sources of ringing, one from the top-hat cutoff and one from \texttt{FFTLog}.}
This is a highly conservative choice --- a smaller number of points in a shorter $k$ span can still achieve low ringing. Since we calculate $\xi_S$ once and interpolate later, we opt for the high accuracy in the \texttt{FFTLog} calculation.

\subsubsection{Spectrograph resolution}
As a further complication, both of these expressions will have a wavelength-dependent spectrograph window function contribution $W(k_\|, \lambda)$. In general, each forest falls onto a different part of the spectrograph and can have a unique spectrograph window function. This is formally treated in ref.~\cite{fontriberaEstimate3DPowerLya2018} using an alternative summary statistics $P_\times(\Delta \theta, k_\|)$. For simplicity, we instead use an average window function of the form:
\begin{equation}
    W(k_\|) = \mathrm{e}^{-k_\|^2 \sigma_\|^2 / 2} \mathrm{sinc}\left(k_\| \Delta r_\| / 2 \right),
\end{equation}
where $\sigma_\| = 69~$\kms$\approx 1~$Mpc and $\Delta r_\| \approx 2.9$ Mpc for eBOSS \lya\ forest data at $z=2.4$ since eBOSS \lya\ data is coarse-grained by a factor of 3.

The spectrograph resolution can be much finer than the mesh grid spacing, so its contributions to the large-scale component can be ignored in the \pcm\ formalism. Still, we use an average window function in the convolution as it does not cost additional computational time. In theory, the small-scale contribution can incorporate unique spectrograph window functions for each quasar by computing $\xi_S(r_\bot, r_\|)$ using $P_S \rightarrow P_S W_a(k_\|) W_b(k_\|)$ for quasars $a$ and $b$ similar to ref.~\cite{fontriberaEstimate3DPowerLya2018}. However, this is extremely costly as it requires the recalculation of $\xi_S(r_\bot, r_\|)$ for every quasar pair.

Lastly, DESI's spectral extraction pipeline provides a resolution matrix $\mathbf{R}_q$ for each quasar \cite{boltonSpectroPerfectionismAlgorithmicFramework2010, guySpectroscopicDataProcessingPipeline2022}. Under the resolution matrix formalism, the spectrograph line smearing can be represented with a block-diagonal matrix $\mathbf{R}$ operating on the entire data vector $\bm d$ such that $\mathbf{S} \rightarrow \mathbf{R} \mathbf{S} \mathbf{R}^\mathrm{T}$. This can be implemented per forest matrix-vector multiplications for both components without modifying eqs.~\eqref{eq:P_LS}~and~\eqref{eq:xi_S_dht} or approximating the window function.


\subsection{Redshift evolution}
The 3D map of the \lya\ forest spans a large volume between $2.0 \leq z \leq 3.8$. The translational symmetry is strongly broken in the line-of-sight direction, and the FFT is not well-defined. We solve this problem by calculating the signal matrix at a pivot redshift $z_\mathrm{pivot}=2.4$ and applying the redshift growth function $G(z)$ separately from the convolution with FFTs.

For a pixel pair $ij$, the redshift-evolved signal matrix is given by $S_{ij} = G(z_{ij}) S^\mathrm{pivot}_{ij}$, where $1 + z_{ij} = \sqrt{(1 + z_i) (1 + z_j)}$. This is an element-wise operation and not trivial in the general case. However, for separable redshift growth functions, $G(z_{ij}) = \sqrt{G(z_i)} \times \sqrt{G(z_j)}$, this element-wise operation becomes
\begin{equation}
    \mathbf{S} = \mathbf{G}^{1/2} \mathbf{S}^\mathrm{pivot} \mathbf{G}^{1/2},
\end{equation}
where $\mathbf{G}^{1/2}$ is a diagonal matrix and now trivial to implement. This is a valid assumption since the redshift growth function will be separable for power law growth functions of $\sim (1 + z_{ij})^\gamma$. We further propagate this growth function formalism to derivative matrices by $\mathbf{C}_{,k} \rightarrow \mathbf{G}^{1/2}\mathbf{C}_{,k}\mathbf{G}^{1/2}$ and estimate \pthreed\ at a known pivot redshift. 

We use $G^{1/2}(z) = b_F(z) D(z) / b_F^\mathrm{pivot} D(z_\mathrm{pivot})$, where $D(z)$ is the linear growth function of the matter field and $b_F(z)$ is the bias of the \lya\ forest. As discussed in section~\ref{sec:model}, $z_\mathrm{pivot} = 2.4$ and $b_F(z) / b_F^\mathrm{pivot} = \left[ (1 + z) / 3.4 \right]^{\alpha_F}$, where $\alpha_F=3.38$.

These simplifying assumptions are sufficient for our work. For more complicated treatments of the redshift evolution, one could use multiple grids at different redshifts and interpolate the results into a single vector. A critical shortcoming of the single-grid formalism is that when metals or DLAs are included in the covariance matrix model, they must be assumed to have the same growth as \lya\ forest.

\subsection{Continuum marginalization}
\label{subsec:continuum}
Another critical challenge in computing ${\bf C}^{-1}{\boldsymbol x}$ arises from the missing mode matrix. This is handled in many 3D correlation function studies by applying a distortion matrix to the theory~\cite{bourbouxCompletedSDSSIVExtended2020}. In this work, we include it as part of the power spectrum estimation.

The missing mode matrix arises because some components in the spectrum are lost at the quasar continuum fitting stage. Usually, $n_C$ continuum modes --- typically polynomials of $\ln \lambda$ --- are fitted out for each quasar, which also removes real signal from the \lya\ forest. We denote these directions in data vector space by ${\boldsymbol u}^{(A)}$, where $A=1,...,n_C N_Q$ and $n_C$ is one plus the order of the continuum fitting polynomial. Typically, $n_C=2$, so for each quasar an overall mean and tilt are projected out. The $A$th component of this vector is non-zero for sightline $\lfloor A/n_C \rfloor$, and zero for the other sightlines. So we have
\begin{equation}
{\boldsymbol\Sigma} = \sigma \sum_{A=1}^{n_C N_Q} 
{\boldsymbol u}^{(A)}{\boldsymbol u}^{(A)\rm T},
~~~\sigma\rightarrow\infty.
\end{equation}
Working with matrices that formally go to $\infty$ is undesirable, so we use the limiting form of the Woodbury formula,
\begin{equation}
({\bf N}+{\boldsymbol\Sigma})^{-1} = {\bf N}^{-1} - \sum_{A,B} [{\bf \Gamma}^{-1}]_{AB} {\boldsymbol t}^{(A)} {\boldsymbol t}^{(B)\rm T},
~~~ \Gamma_{AB} = {\boldsymbol u}^{(A)\rm T}{\bf N}^{-1}{\boldsymbol u}^{(B)},
~~~
{\boldsymbol t}^{(A)} = {\bf N}^{-1}{\boldsymbol u}^{(A)},
\label{eq:Woodbury}
\end{equation}
which no longer involves infinite entries.
Since ${\bf N}$ is diagonal, the matrix ${\bf \Gamma}$ breaks into $N_Q$ blocks of size $n_C\times n_C$ each, and $({\bf N}+{\boldsymbol\Sigma})^{-1}$ breaks into $N_Q$ blocks of size $n_\lambda\times n_\lambda$ each. We can further define $\bm{\tilde{u}}^{(A)} = \mathbf{N}^{-1/2} \bm{u}^{(A)}$ to write $\Gamma_{AB} = \bm{\tilde{u}}^{(A)\rm T}\cdot \bm{\tilde{u}}^{(B)}$ and
\begin{equation}
    (\mathbf{N} + \mathbf{\Sigma})^{-1} = \mathbf{N}^{-1/2} \left( \mathbf{I} - \sum_{A,B} [{\bf \Gamma}^{-1}]_{AB} \bm{\tilde{u}}^{(A)} \bm{\tilde{u}}^{(B)\rm T} \right) \mathbf{N}^{-1/2}.
\end{equation}
There are several paths forward to consider here. We find that the square root of the matrix in parentheses is easier to implement:
\begin{equation}
     \mathbf{I} - \sum_{A,B} [{\bf \Gamma}^{-1}]_{AB} \bm{\tilde{u}}^{(A)} \bm{\tilde{u}}^{(B)\rm T} = \mathbf{\Pi}^{1/2} \mathbf{\Pi}^{1/2},
\end{equation}
such that $\left(\mathbf{N} + \mathbf{\Sigma}\right)^{-1} = \mathbf{N}^{-1/2}\mathbf{\Pi}^{1/2} \mathbf{\Pi}^{1/2} \mathbf{N}^{-1/2}$. We compute $\mathbf{\Pi}^{1/2}$ once using eigenvalue decomposition for each quasar. These matrices are stored in files and read when needed, as they significantly increase the memory usage.

The covariance matrix-vector multiplication $\mathbf{C}\bm y = (\mathbf{S} + \mathbf{N} + \mathbf{\Sigma}) \bm y = \bm x$ can be written as
\begin{equation}
    (\mathbf{S} + \mathbf{N}^{1/2}\mathbf{\Pi}^{-1/2} \mathbf{\Pi}^{-1/2} \mathbf{N}^{1/2}) \bm y = \bm x
\end{equation}
or
\begin{equation}
    \label{eq:cgd_cm}(\mathbf{I} + \mathbf{\Pi}^{1/2} \mathbf{N}^{-1/2} \mathbf{S} \mathbf{N}^{-1/2} \mathbf{\Pi}^{1/2}) \mathbf{\Pi}^{-1/2} \mathbf{N}^{1/2} \bm y = \mathbf{\Pi}^{1/2}\mathbf{N}^{-1/2} \bm x,
\end{equation}
such that $\bm m = \mathbf{\Pi}^{1/2}\mathbf{N}^{-1/2} \bm x$, and $\bm y = \mathbf{N}^{-1/2} \mathbf{\Pi}^{1/2}\bm z$. As an additional advantage of this formalism, random noise realizations in MC simulations remain well-defined as matrix multiplications on finite values of $\mathbf{N}^{-1/2} \bm n \sim \mathcal{N}(\mu=0, \sigma=1)$.

\subsection{Multiplication with derivative matrices}
Another core operation required by the optimal estimator is of the form $\bm x^\mathrm{T} \mathbf{C}_{, k} \bm x$. As noted in the previous section, the derivative matrix includes a redshift evolution $\mathbf{C}_{,k} = \mathbf{G}^{1/2}\mathbf{C}_{,k}^\mathrm{pivot} \mathbf{G}^{1/2}$. Therefore, devolving redshift evolution from the vector $\bm {x}$ entails a simple $\bm x \rightarrow \mathbf{G}^{1/2} \bm x$ operation. The \pcm\ methodology still allows us to divide this operation into large- and small-scale components: $\bm x^\mathrm{T} (\mathbf{C}_{L, k} + \mathbf{C}_{S, k}) \bm x$.

The power spectrum can be estimated as $P(k_\bot, k_\|)$ or in multipoles $P_\ell(k)$ such that $P(k, \mu) = \sum_{\ell=0}^{\ell_\mathrm{max}} P_\ell(k) L_\ell(\mu)$ where $L_\ell$ are the Legendre polynomials. While we describe both approaches below, our analysis focuses on estimating multipoles.

\subsubsection{Measuring \texorpdfstring{$P(k_\bot, k_\|)$}{P(k\_perp, k\_parallel)}}
In Fourier space, $\mathbf{\tilde C}_{L, k}$ is a diagonal matrix where the diagonals are $\exp(-k^2 / R_L^2) W^2(k_\|)$ if the Fourier mode is in $(k_\bot, k_\|)$ bin, and zero otherwise. Therefore, $\bm {\tilde x}^\dagger \mathbf{\tilde C}_{L, k} \bm {\tilde x}$ is a simple calculation that can be done purely on the mesh. The small-scale derivative matrices can be constructed using the derivative of eq.~\eqref{eq:xi_S_dht}:
\begin{align}
    \mathbf{C}_{S, k}(r_\bot, r_\|) &= \int_{k_\bot^{(1)}}^{k_\bot^{(2)}} \frac{k_\bot \mathrm{d}k_\bot}{2 \pi} J_0(k_\bot r_\bot) \int_{k_\|^{(1)}}^{k_\|^{(2)}} \frac{\mathrm{d}k_\|}{\pi} \cos(k_\| r_\|) W^2(k_\|) (1 - e^{-k^2/r^2_L}) \nonumber \\
    &= q^{T, k}_\bot(r_\bot) q^{T, k}_\|(r_\|) - q^{L, k}_\bot(r_\bot) q^{L, k}_\|(r_\|),
\end{align}
where superscripts (1) and (2) denote the lower and upper bin edges, respectively, and we completely separated $k_\bot$ and $k_\|$ integrations and defined the integrals:
\begin{align}
    q^{T, k}_\bot(r_\bot) &= \int_{k_\bot^{(1)}}^{k_\bot^{(2)}} \frac{k_\bot \mathrm{d}k_\bot}{2 \pi} J_0(k_\bot r_\bot),& 
    q^{L, k}_\bot(r_\bot) &= \int_{k_\bot^{(1)}}^{k_\bot^{(2)}} \frac{k_\bot \mathrm{d}k_\bot}{2 \pi} J_0(k_\bot r_\bot) e^{-k_\bot^2/r^2_L}, \nonumber\\
    q^{T, k}_\|(r_\|) &= \int_{k_\|^{(1)}}^{k_\|^{(2)}} \frac{\mathrm{d}k_\|}{\pi} \cos(k_\| r_\|) W^2(k_\|) ,&{\rm and~} q^{L, k}_\|(r_\|) &= \int_{k_\|^{(1)}}^{k_\|^{(2)}} \frac{\mathrm{d}k_\|}{\pi} \cos(k_\| r_\|) W^2(k_\|) e^{-k_\|^2/r^2_L}.
\end{align}
The $q^{T, k}_\bot$ integration can be done exactly:
\begin{align}
    q^{T, k}_\bot(r_\bot) &= \frac{k_\bot^{(2)} J_1(k_\bot^{(2)} r_\bot) - k_\bot^{(1)} J_1(k_\bot^{(1)} r_\bot)}{2\pi r_\bot};& \mathrm{note}\, \lim_{r_\bot \to 0} q^{T, k}_\bot(r_\bot) = \frac{k_{\bot, (2)}^2 - k_{\bot, (1)}^2}{4\pi}.
\end{align}

\subsubsection{Measuring multipoles \texorpdfstring{$P_\ell(k)$}{P\_ell(k)}}
In practice, most of the \pthreed\ information from the \lya\ forest alone (based on the best-fitting values to the model of ref.~\cite{arinyoNonLinearPowerLya2015}) is contained in the $\ell=0, 2, 4$ multipoles at linear scales \cite{kirkbyFittingMethodsBaryon2013, belsunce3dLymanAlphaPowerSpectru2024}.
However, metal lines exhibit distinct oscillation signatures in $\ell \geq 6$ multipoles, so we also estimate higher-order multipoles. The reduced dimensionality of the multipole approach greatly helps the precision of the estimated covariance matrix.

In Fourier space, $\mathbf{\tilde C}_{L, k}$ remains a diagonal matrix, with diagonal elements given by $\exp(-k^2 / R_L^2)$ $ W^2(k_\|) L_\ell(\mu)$ if the Fourier mode falls within a $k$ bin, and zero otherwise. This introduces an additional $L_\ell(\mu)$ weighting compared to Cartesian binning. We apply linear interpolation across $k$ bins in estimating multipoles, introducing another term --- triangular binning function \cite{karacayliOptimal1DLy2020}. This correlates neighboring $k$ bins, but as shown in section~\ref{sec:validation}, the bins are already fairly correlated.

Small-scale corrections can be derived following the steps above. However, we ignore these contributions here since the measurement precision does not necessitate their inclusion.

\subsection{Estimating the bias and the Fisher matrix\label{subsec:estimate_bias_fisher}}
We estimate the bias term $b_k = \mathrm{Tr}(\mathbf{C}^{-1}\mathbf{C}_{,k})$ using the stocastic trace estimation method. We generate Monte Carlo (MC) realizations of the data vector $\bm q$, such that $\langle \bm q \bm q^\mathrm{T} \rangle = \mathbf{C}$, and
\begin{equation}
    b_k = \mathrm{Tr}(\mathbf{C}^{-1}\mathbf{C}_{,k} \mathbf{C}^{-1} \mathbf{C}) = \langle \bm q^\mathrm{T} \mathbf{C}^{-1}\mathbf{C}_{,k} \mathbf{C}^{-1} \bm q \rangle.
\end{equation}
As noted in Oh~et~al.~\cite{ohEfficientPowerSpectrumCmb1999}, these MC simulations give the Fisher matrix for free:
\begin{equation}
    {F}_{kk'} = \frac{1}{4} \left\langle (b_k - \langle b_k \rangle) (b_{k'} - \langle b_{k'} \rangle) \right\rangle.
\end{equation}
We find this to be the fastest method to obtain the Fisher matrix, albeit $N_\mathrm{MC} \gtrsim 2,000$ realizations are required to suppress noise in the off-diagonal terms. We use the estimated variance on the mean to quantify the accuracy of our MC simulations $\bm b^{(i)}$, which is given by $\sum_i (\bm b^{(i)} - \langle \bm b \rangle )^2 / N_\mathrm{MC} (N_\mathrm{MC} - 1)$, where $\langle \bm b \rangle = \sum_i \bm b^{(i)} / N_\mathrm{MC}$. 
After 2,880 realizations, we reach an average $6 \times 10^{-5}$ deviation. 

The primary challenge in this front arises from incorporating correlated small-scale fluctuations between sightlines. To address this, we construct a composite random vector \( \bm{q} = \bm{q}_L + \bm{q}_S \), where the two components are independent and have covariances $\langle \bm q_L \bm q_L^{\rm T} \rangle = {\bf C}_L \equiv {\bf S}_L$ and $\langle \bm q_S \bm q_S^{\rm T} \rangle = {\bf C}_S \equiv {\bf S}_S+{\bf N}$ respectively, where ${\bf S}_L$ and ${\bf S}_S$ are the long-range (PM) and short-range (PP) signal matrices. Theoretically one should include a $\bm q_\Sigma$ with covariance ${\boldsymbol\Sigma}$ as well, which would be a linear combination of continuum modes with formally infinite variance; but in the limit $\sigma\rightarrow\infty$, the covariance of ${\bf C}^{-1}\bm q_\Sigma$ goes to zero, so it is not necessary to include $\bm q_\Sigma$.

The large-scale component \( \bm{q}_L \) is generated via convolution on the mesh by generating white noise in real space and convolving with $\sqrt{P_L(\bm k)/\Delta V}$ where $\Delta V$ is the cell volume, and interpolated trilinearly to forest pixel locations. 

The large-scale field obtained through convolution on the mesh lacks the small-scale correlations and pipeline noise fluctuations, given by $\mathbf{A}_S = \mathbf{I} + \mathbf{N}^{-1/2} \mathbf{S}_S \mathbf{N}^{-1/2}$. To incorporate these fluctuations, we first generate unit Gaussian fields at the forest pixel locations. Denoting this vector as \(\bm \eta \), we obtain \(\bm{q}_S = {\bf N}^{1/2} \mathbf{A}_S^{1/2} \bm \eta \), which has the desired covariance. For numerical reasons, ${\bf N}^{-1/2}\bm q$ is stored for both cases. 
This requires us to implement the square root of a matrix (${\bf A}_S^{1/2}$) operating on a vector (${\boldsymbol\eta}$). There are various iterative algorithms to compute the matrix square root  $\mathbf{Q} = \mathbf{A}_S^{1/2}$, mainly based on Newton's method and the matrix sign function \cite{schulzIterativeMatrix1933, KennyLaubRationalIterativeMatrix1991, pandeyParalelAlgorithmMatrixSign1990, KennyLaubHyperbolicPade1994, highamStableIterationsMatrixSquareRoot1997, petkovicMatrixSquareRoot1998}. In appendix~\ref{app:matrix_sqrt}, we review these methods. Among them, the single-iteration Padé approximation of order $p$ demonstrates the best performance:
\begin{equation}
    \mathbf{Q}_p = \frac{\mathbf{A}_S}{p} \sum_{i=1}^p \frac{1}{\xi_i} (\mathbf{A}_S  + \alpha_i^2\mathbf{I})^{-1},~~ \mathrm{where~} \xi_i= \frac{1}{2}\left(1 + \cos\frac{(2i-1)\pi}{2p}\right)~\mathrm{and~} \alpha_i^2=\frac{1}{\xi_i} - 1,
\end{equation}
where we first shrink the matrix by the largest diagonal, $\mathbf{A}_S \rightarrow \mathbf{A}_S/\mathrm{max}(A_{S, ii})$, and later scale up the results by $\sqrt{\mathrm{max}(A_{S, ii})}$. While computing $p$ matrix inverses may initially appear excessively costly, their convergence is accelerated by the addition of $\alpha_i^2$ to the diagonal. Specifically, for $p=4$, we find all four matrix inverses converge in 42 iterations in total for $\epsilon=10^{-5}$.

\begin{figure}
    \centering
    \includegraphics[width=0.46\linewidth]{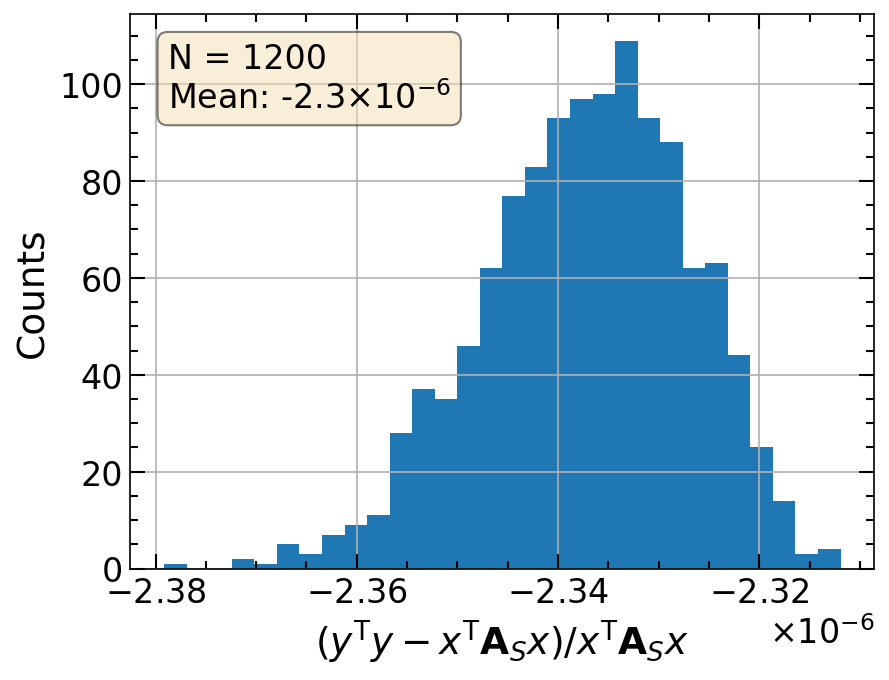}
    \quad
    \includegraphics[width=0.46\linewidth]{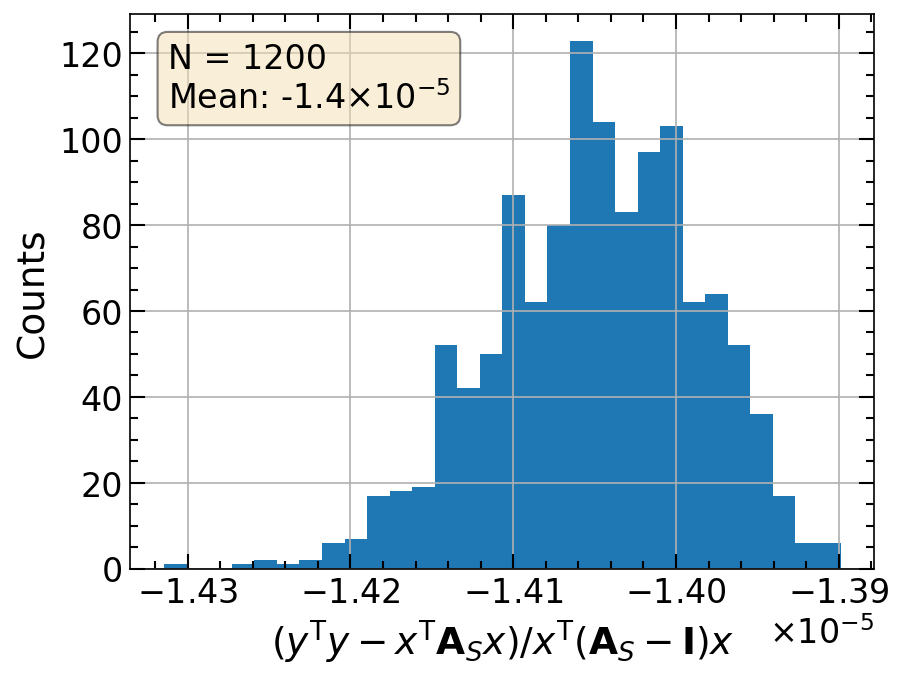}
    \caption{({\it Left}) A histogram of relative errors in the square root of $\mathbf{A}_S$ estimate, where $\bm y \equiv \mathbf{Q}_4 \bm{x}$. From 1,200 random vectors $\bm x$, the mean relative error indicates a $-2.3\times 10^{-6}$ bias. ({\it Right}) To assess how well the small-scale signal is captured, we subtract the identity matrix from $\mathbf{A}_S$ and calculate relative errors with respect to $\mathbf{A_S} - \mathbf{I}$. The mean relative error bias remains small at $-1.4\times 10^{-5}$.}
    \label{fig:hsqrt_relerr}
\end{figure}

To test for the accuracy of this procedure, we draw 1,200 random vectors $\bm x$, calculate the dot product $\bm{y}^\mathrm{T} \bm{y}$, where $\bm y \equiv \mathbf{Q}_4 \bm{x}$, and compare it to true dot product $\bm x^\mathrm{T} \mathbf{A}_S \bm x$. Figure~\ref{fig:hsqrt_relerr} shows a histogram of the relative error on this dot product.
We observe the fourth order to be sufficient with a small mean bias of approximately $10^{-6}$.
When assessing how well the small-scale signal is captured by removing ones from the diagonal, the mean relative bias slightly increases to approximately \( 10^{-5} \). The fifth-order approximation reduces these biases by a factor of 30, but converges in 52 iterations, resulting in a 22\% increase in computational cost.

Alternatively, one can define a ``noise bias'' similar to \poned\ formalism $b^N_k = \mathrm{Tr} ( \mathbf{C}^{-1} \mathbf{C}_{,k} $ $\mathbf{C}^{-1} \mathbf{N})$,
in which case the MC simulations only draw random noise vectors $\bm n$ without a need for any signal component, and the quadratic estimator in eq.~\eqref{eq:p3d_estimator} will not have the fiducial \pthreed\ term:
\begin{equation}
    \hat P^\mathrm{3D}_{k} = \sum_{k'} \frac{1}{2} F^{-1}_{kk'}\left[\bm{d}^\mathrm{T} \mathbf{C}^{-1}\mathbf{C}_{,k'} \mathbf{C}^{-1} \bm{d} -  b^N_{k'} \right].
\end{equation}
However, the Fisher matrix cannot be obtained from these simulations. The same disadvantage applies to other stochastic estimation methods that do not draw from the covariance, such as a vector with random entries of $\pm 1$ used in ref.~\cite{padmanabhanMiningWeakLensing2003}.

There is a second, but slower method for the Fisher matrix estimation: direct calculation of the trace $F_{kk'} = \langle \bm \eta^\mathrm{T} \mathbf{C}^{-1} \mathbf{C}_{,k} \mathbf{C}^{-1}  \mathbf{C}_{,k'} \bm\eta \rangle$ by generating random vectors $\bm\eta$ of $\pm 1$s.
This method yields less noisy Fisher matrix estimates even with $\sim 500$ random vectors, but requires $N_k$ conjugate gradients and two additional FFTs for every random vector $\bm\eta$, which becomes unfeasibly slow unless some time-saving measures are taken. Using the preconditioner (block-diagonal approximation) instead of the conjugate gradient method makes this estimation achievable but fails to capture correlations between multipoles and $k$ bins. 


Lastly, we briefly describe some of our unsuccessful attempts for future reference. One avenue we pursued was based on transforming the equation for the bias to $b_k = \langle \bm{q'}^{\mathrm{T}} \mathbf{C}_{,k} \bm{q'} \rangle$, where $\bm{q'}$ is drawn from $\mathcal{N}(\bm 0, \mathbf{C}^{-1})$. There are many algorithms to draw samples from a high-dimensional Gaussian covariance matrix (see ref.~\cite{vonoHighDimensionalGaussianSampling2020} for an overview). The conjugate gradient solver can be transformed to a conjugate gradient sampler with minimal change~\cite{parkerSamplingGaussianDistributions2012}.
However, as noted in refs.~\cite{vonoHighDimensionalGaussianSampling2020} and \cite{parkerSamplingGaussianDistributions2012}, this method is only suitable if eigenvalues are well-separated. This is especially problematic for $\mathbf{C}^{-1}$ in our current formalism since the solver prefers the condition number to be small. The references further note that the performance of the conjugate gradient sampler significantly degrades for vectors drawn from $\mathbf{C}^{-1}$. Other methods might prove more fruitful. We also attempted to draw random vectors from $\mathbf{C}_{,k}$ to estimate the Fisher matrix directly. This operation requires the square root of the derivative matrix and can be done in Fourier space for the $(k_\bot, k_\|)$ binning scheme. However, the multipole estimation introduces negative values to Fourier modes and cannot be done without supporting imaginary numbers.

\subsection{Some numerical details}
Our initial tests show that shared-memory parallelization through \texttt{OpenMP} saturates at 8--16 cores.
Most high-performance computing clusters have more cores and have enough memory to run more than one job in a single node. Therefore, we simultaneously run multiple MC simulations by applying \texttt{MPI} parallelization along with \texttt{OpenMP}. The synchronization between \texttt{MPI} tasks is not required, so it can be kept at a minimum level. For example, the CCAPP Condo on The Ohio Supercomputer Center (OSC) \cite{OhioSupercomputerCenter1987}, where we carried out our computations, has 96 cores per compute node. We typically run the estimator using 12 MPI tasks, each utilizing 8 \texttt{OpenMP} threads per node.

The integrals and complicated terms such as $P_L$, \poned,  and \pthreed\ are interpolated on a regular grid that eliminates binary search and is preferable over a general grid implementation. In 1D interpolation, a cubic spline offers smoothness and incurs minimal evaluation time. In 2D interpolation for $\xi_S(r_\bot, r_\|)$, cubic spline becomes costly, but the fastest method, linear interpolation, compromises smoothness.
Cubic Hermite spline interpolation achieves smoothness without a hefty computational cost. Specifically, we employ the Bernstein polynomial \cite{bernstein1912} formulation of cubic Hermite interpolation and evaluate it using de Casteljau's recursive algorithm \cite{deCasteljau1959outillages} (see appendix~\ref{app:hermite} and ref.~\cite{farin2001curves} for details).

There is one last computational optimization we apply to correlation function interpolations. $\xi_S(r_\bot, r_\|)$ and $\xi_{1\mathrm{D}}(r_\|)$ are smoother and slowly changing in $\log r$. However, evaluating $\log r$ for every pixel pair is unexpectedly expensive. Given that double-precision accuracy for $\log r$ is unnecessary, we employ a bitwise approximation of $\log_2 r$,\footnote{\url{https://github.com/romeric/fastapprox}} achieving nearly an order of magnitude improvement in evaluation speed.

\section{Validation\label{sec:validation}}
We test our estimator on flat-sky mocks. Aside from the simplified geometry, we want to work with a mock catalog relevant to present surveys and, therefore, build a mock catalog with background quasar coordinates and forest pixel locations as well as pixel noise properties drawn from a real survey. 

We focus on the value-added \lya\ forest catalog, which contains transmitted flux fluctuations $\delta_F$, from the 16th data release (DR16) of the complete extended Baryonic Oscillation Spectroscopic Survey (eBOSS) \cite{bourbouxCompletedSDSSIVExtended2020}. The dataset includes 210,005 quasars in the redshift range $2.1 \lesssim z_\mathrm{qso} \lesssim 4.0$, with nearly 85\% at $z_\mathrm{qso} < 3.0$ in this dataset. The large-scale structure fluctuations $\sigma^2_\mathrm{LSS}$ are included in the weights in this value-added catalog. Our method requires pipeline noise estimates directly, so we subtract the $\sigma^2_\mathrm{LSS}$ contribution from the weights.
The Galactic plane divides the data into two regions: the south Galactic cap (SGC) contains 58,592 quasars, while the north Galactic cap (NGC) holds the remaining 151,413.

We find that the data in SGC provides enough statistics to study our algorithm, which requires about 30 node-hours with the configuration detailed below. Although our formalism is efficient, it remains expensive, and testing it with the entire sample increases the cost by approximately a factor of three with minimal return. So, our analysis is restricted to the SGC within the redshift range $2.1\leq z_\mathrm{Ly\alpha} \leq 2.7$, corresponding to an effective radial distance of 5.8 Gpc and Cartesian box dimensions of (13.2 Gpc, 4.97 Gpc, 755 Mpc). Using a mesh grid of (2048, 768, 384), the transverse grid spacing is 6.5 Mpc. To realize uniform grid spacing across all dimensions, we extend the radial length to 2.48 Gpc by padding both ends along the line-of-sight direction.

Based on these parameters, we set $R_L=6~$Mpc and $f=8$ as our baseline. Each forest has an average of 10.3 neighboring forests within a 48 Mpc radius.  We estimate $P_\ell(k)$ for $l=0\dots8$ in 20 linearly-spaced bins at $k>0.015~$\impc, with $\Delta k=1.2\times 10^{-2}~$\impc, a scale that is several times larger than the fundamental frequency in the line-of-sight direction.

We set the convergence criteria to $\epsilon=10^{-6}$ and run 2,880 MC simulations to estimate the Fisher matrix.


\subsection{Self-consistency of Monte Carlo simulations}
We generate mock \lya\ data using our Monte Carlo (MC) simulation procedure to demonstrate that: (1) our method produces correlated sightlines with the correct \pthreed; and (2) the estimator successfully recovers the input \pthreed.

\subsubsection{Simulations with only \texorpdfstring{\lya}{Lya} forest}
We start by only considering the \lya\ forest in the signal. The conjugate gradient method converges in 40 iterations on average.

\begin{figure}
    \centering
    \includegraphics[width=0.47\linewidth]{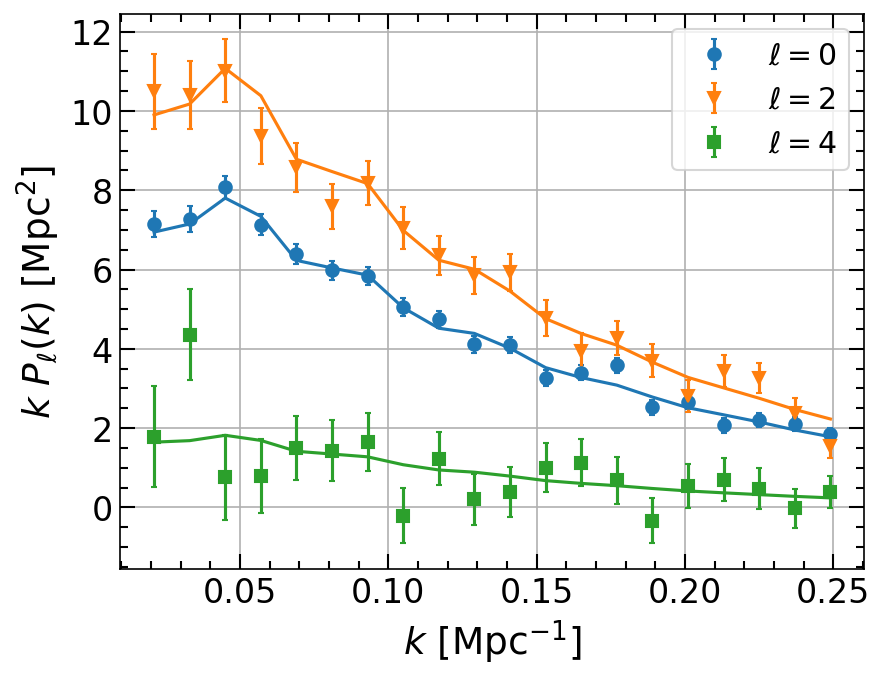}
    \hfill
    \includegraphics[width=0.47\linewidth]{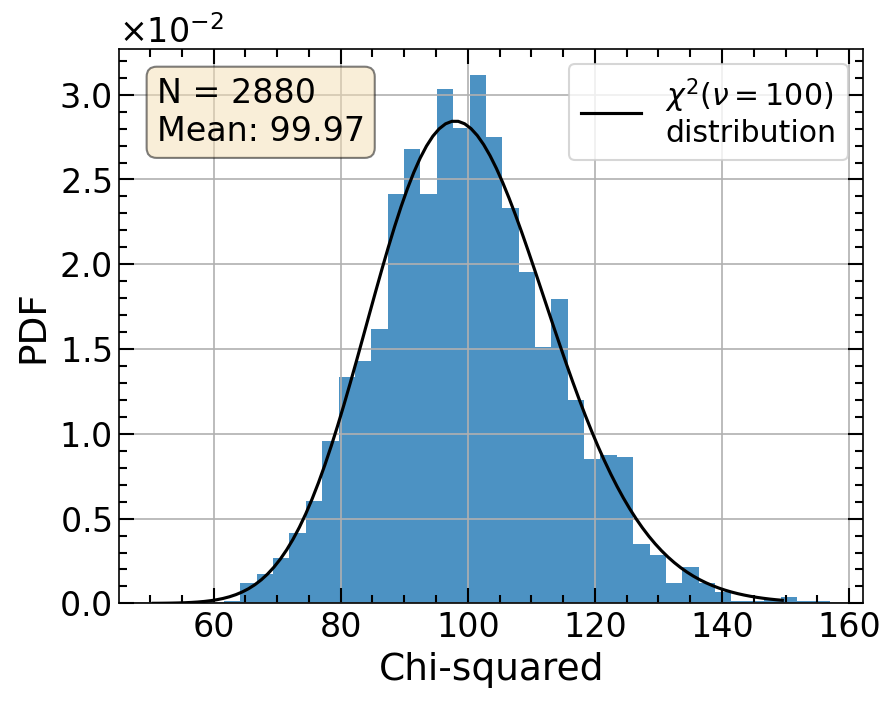}
    \caption{({\it Left}) \pthreed\ multipole estimations in SGC from one MC realization without any contaminants. \lya\ forest quadrupole is stronger than its monopole. $l=6$ multipole is effectively zero and not shown for clarity. ({\it Right}) Histogram of $\chi^2$s from 2,880 MC realizations. The probability distribution function follows the theoretically expected $\chi^2(\nu = 100)$ distribution.}
    \label{fig:p3d_est_vanilla}
\end{figure}

Figure~\ref{fig:p3d_est_vanilla} shows multipole estimates from one mock data set on the left panel. All 2,880 MC realizations are independent mock data by themselves. Since these are Gaussian, the calculated $\chi^2$ follows the theoretically expected $\chi^2(\nu = 100)$ distribution, which can be seen in the right panel.

\begin{figure}
    \centering
    \includegraphics[width=0.95\linewidth]{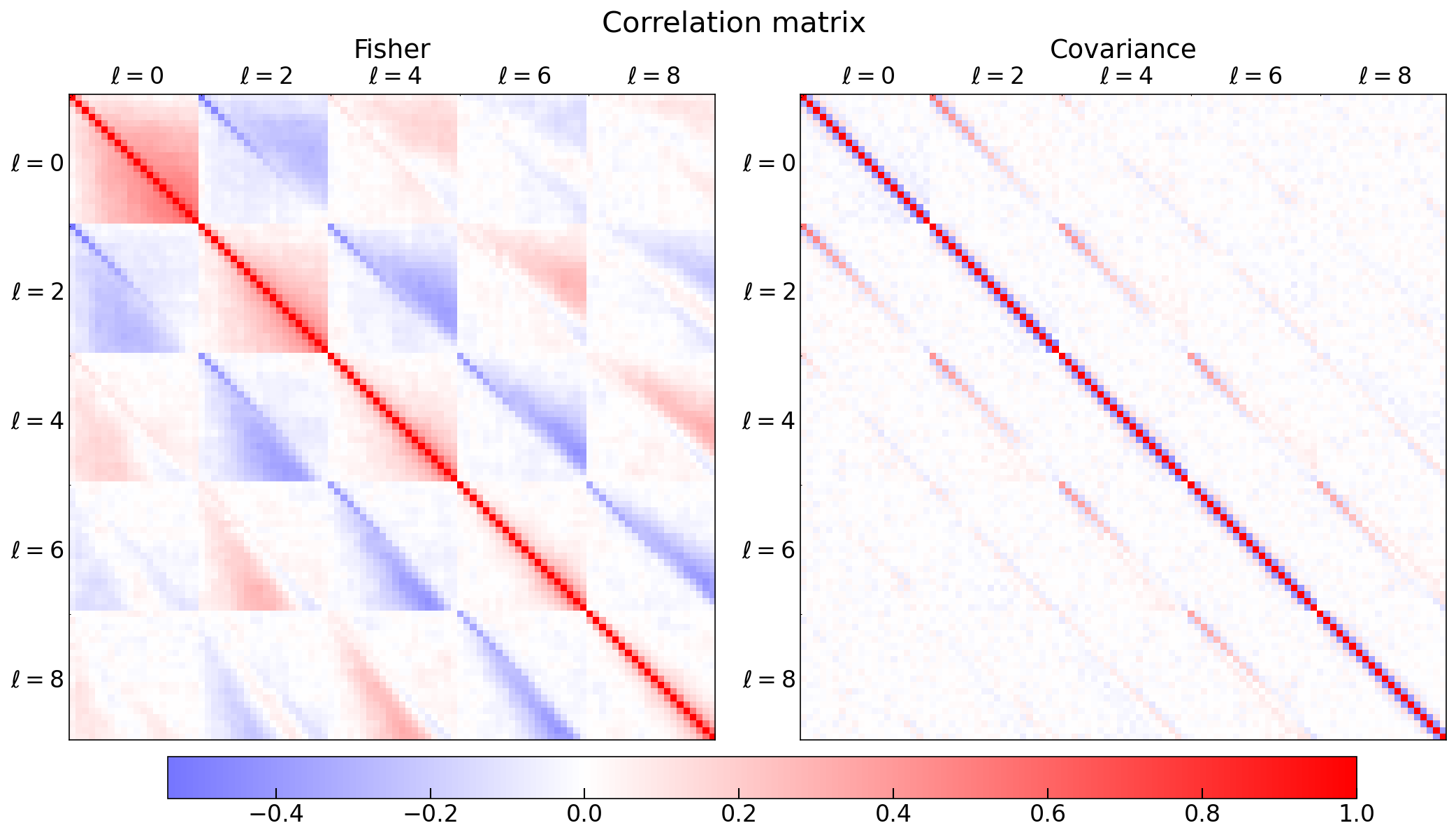}
    \caption{Fisher correlation matrix on the left and correlation matrix between multipoles on the right. Correlations between $k$ bins and multipoles seem weak in the covariance matrix, but the Fisher matrix manifests stronger correlations. These should be taken into account in likelihood calculations for cosmological inference.}
    \label{fig:fisher_cov_vanilla.png}
\end{figure}

Figure~\ref{fig:fisher_cov_vanilla.png} shows the correlation matrices for the Fisher matrix on the left and the covariance matrix on the right. The neighboring $k$ bins are correlated due to linear interpolation, but --- as we alluded to in the previous section --- all $k$ bins are correlated, especially bins within the monopole. Correlations between $k$ bins and multipoles seem weak in the covariance matrix, but the inverse (Fisher) matrix manifests stronger correlations. Since the likelihood calculations are based on the inverse, it is essential to consider these correlations.

\subsubsection{Including HCD and metal systems}
We now demonstrate how our prescription for HCDs and ionized silicon systems alters the measurement, with contaminant model parameters noted in section~\ref{sec:model}. The conjugate gradient method converges in 50 iterations on average --- a large increase from the previous case.

\begin{figure}
    \centering
    \includegraphics[width=0.95\linewidth]{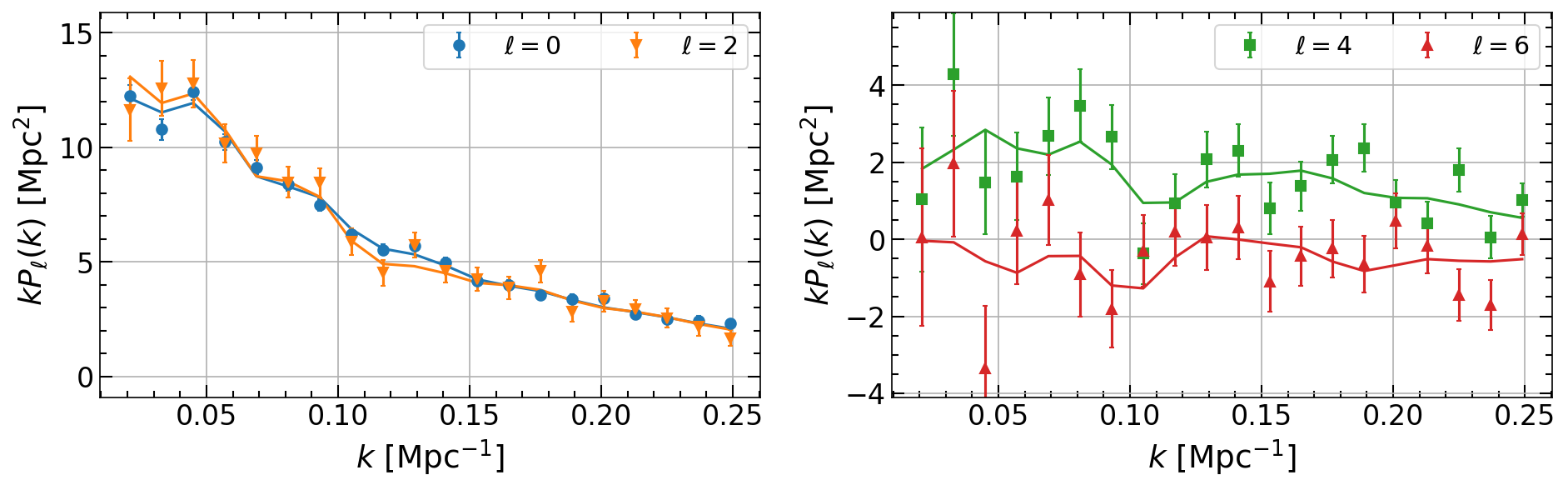}
    \caption{Measured \pthreed\ multipoles in SGC from one MC realization with HCD and ionized silicon contaminants. The monopole strengthens relative to the quadrupole due to contributions from contaminants. The oscillations induced by Si~\textsc{ii} and Si~\textsc{iii} generate a non-negligible $l=6$ signal, though the constraining power of SGC forests is insufficient for decisive detection.}
    \label{fig:p3d_est_wconta}
\end{figure}

Figure~\ref{fig:p3d_est_wconta} shows the measured \pthreed\ multipoles in SGC from one mock data set with contaminants.
Due to additional power from HCD and ionized silicon contaminants, the monopole term is now equal to the quadrupole term. As noted in previous sections, the redshift confusion of ionized silicon systems yields a non-negligible $l=6$ signal via oscillations in the power spectrum. The SGC quasars alone are not enough to decisively detect $P_{6}$.

\subsubsection{Continuum marginalization}
A pillar of optimal estimators is the marginalization of contaminated modes --- in our case, quasar continuum fitting errors. We account for typical errors introduced by the eBOSS and DESI pipelines, modeled as amplitude and slope terms for each forest \cite{bourbouxCompletedSDSSIVExtended2020, ramirezperezLyaCatalogDesiEdr2023}. However, continuum error templates can be adapted for novel approaches, including machine learning and neural network-based methods \cite{sunQuasarFactorAnalysis2023, turnerLyaForestMeanFluxFromDesiY12024}.

Our algorithm requires eigenvalue decomposition for each forest, along with storing and retrieving a projection matrix for every quasar from the disk --- making implementation the primary challenge rather than effectiveness. Here, we focus on the implementation and its overall impact, while the efficacy of the method is discussed in section~\ref{subsubsec:continuum}.

We assess the computational feasibility and convergence behavior of the conjugate gradient method using the same setup as before, but without any contamination or added continuum errors. Each multiplication with the matrix $\mathbf{A}$ requires two disk accesses per forest. Despite this significant requirement, reading the matrix $\mathbf{\Pi}$ and multiplying it with vectors constitute only 3.3\% of the total computational time for the $\mathbf{A} \bm{z}$ operation. This efficiency is thanks to OSC’s high-performance disk space per compute node and our use of raw binary files, which can be read serially by each \texttt{OpenMP} thread. 



\begin{figure}
    \centering
    \includegraphics[width=0.45\linewidth]{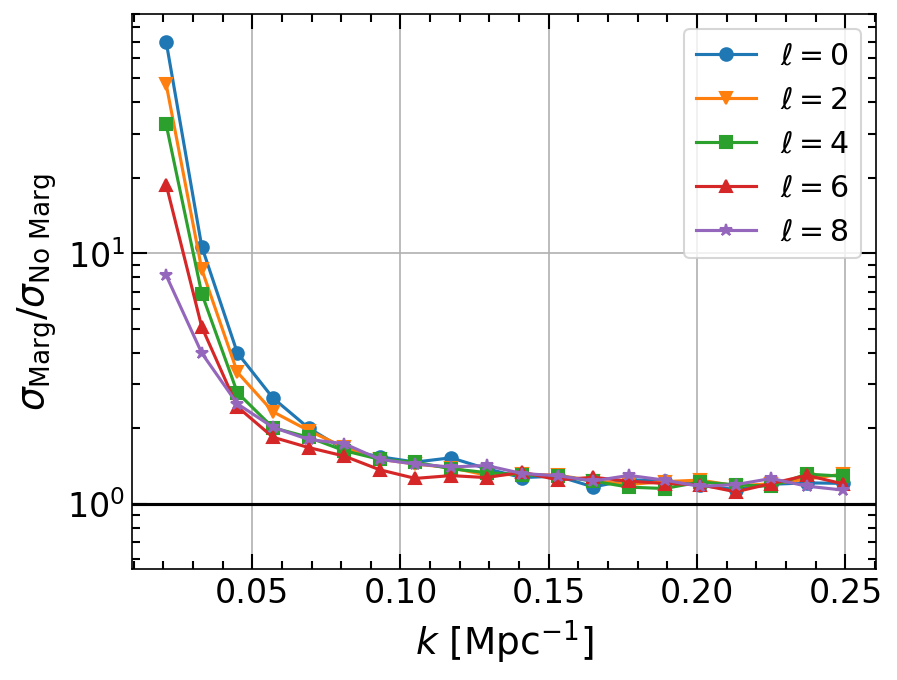}
    \hfill
    \includegraphics[width=0.52\linewidth]{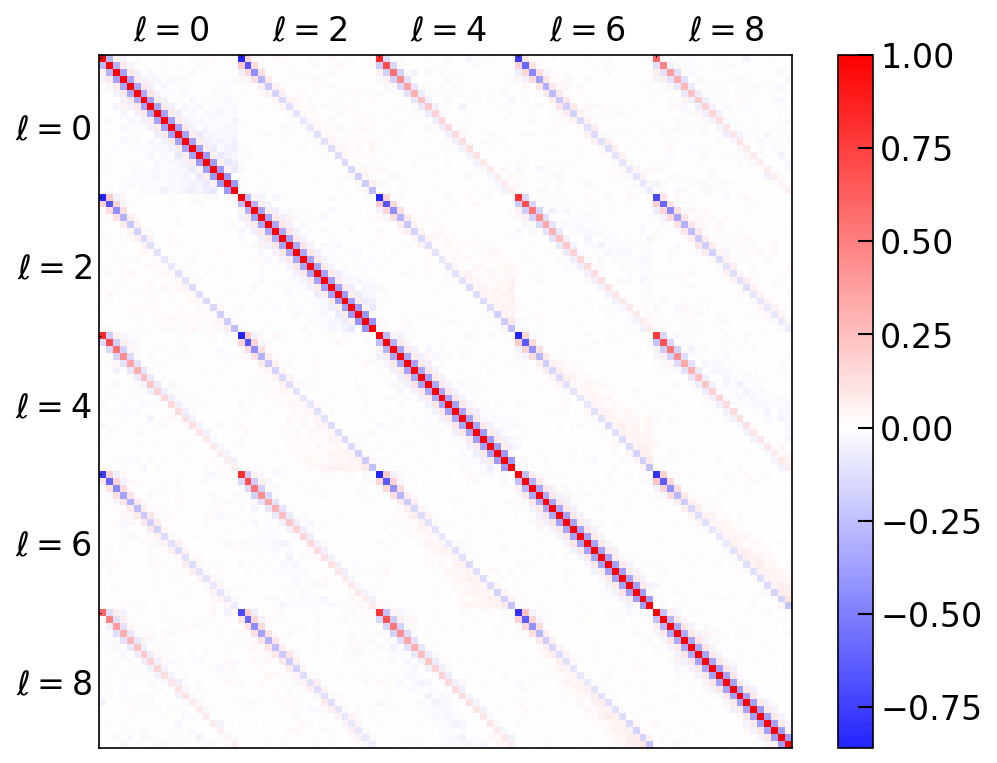}
    \caption{({\it Left}) Increase in error after continuum marginalization. Errors in low-$k$ modes are significantly inflated. ({\it Right}) Correlation matrix between multipoles in the continuum marginalization case. In contrast to the unmarginalized case, the positive correlations between $\delta \ell = 2$ modes transform to anti-correlations, and correlations extend to higher $\delta \ell$.}
    \label{fig:continuum_marg_case}
\end{figure}

The effect of continuum marginalization on \pthreed\ mirrors its effect on \poned: it inflates errors in low-$k$ modes. Specifically, this negative impact would have been localized to low $k_\|$ modes in the $(k_\bot, k_\|)$ binning scheme \cite{blomqvistBroadbandDistortionLya2015}. In the multipole expansion, the effect of marginalization on these modes spreads over a range of $k$ bins. This can be seen on the left panel of figure~\ref{fig:continuum_marg_case}.
For the lowest bin at $k = 2.1 \times 10^{-2}~$\impc, the errors increase by a factor of nearly $35$ across all multipoles. For the next bin at $k=3.3\times 10^{-2}~$\impc, the errors increase by a factor of $7$ compared to the case without continuum marginalization. For modes with $k \gtrsim 5.0 \times 10^{-2}~$\impc, the errors are 30\% larger on average.

Additionally, the correlations between multipoles are amplified due to the marginalized modes. The right panel of figure~\ref{fig:continuum_marg_case} illustrates the correlation coefficient derived from the covariance matrix. In contrast to the unmarginalized case, the positive correlations between $\delta \ell = 2$ modes transition to anti-correlations, and correlations generally extend to $\delta \ell = 8$ and, most likely, even larger $\delta \ell$s.

\subsection{Gaussian simulations}
In this section, we evaluate the validity of our formalism using an externally generated data vector. Preliminary analysis based on eBOSS data and mocks revealed that projections from spherical to Cartesian geometry introduce significant distortions in the measurements. Therefore, we restrict our external data set to cases without spherical geometry and defer a comprehensive treatment of spherical geometry to future work.

We generate a random Gaussian field on a Cartesian grid with three times finer spacing, convolved with the total \lya\ power spectrum, and interpolate this field to the eBOSS \lya\ forest data locations. We then add Gaussian noise consistent with the pipeline to the synthetic data. This approach captures the eBOSS survey window function without introducing distortions.

As discussed in section~\ref{sec:model}, \pthreed\ extends to smaller angular scales compared to radial scales, as they are constrained only by the pressure-smoothing scale. Without appropriate suppression, the angular modes alias in a grid of 2.2~Mpc spacing. To prevent this problem, we adopted an additional isotropic suppression scale of $k_A=0.4$~\impc.

\begin{figure}
    \centering
    \includegraphics[width=0.47\linewidth]{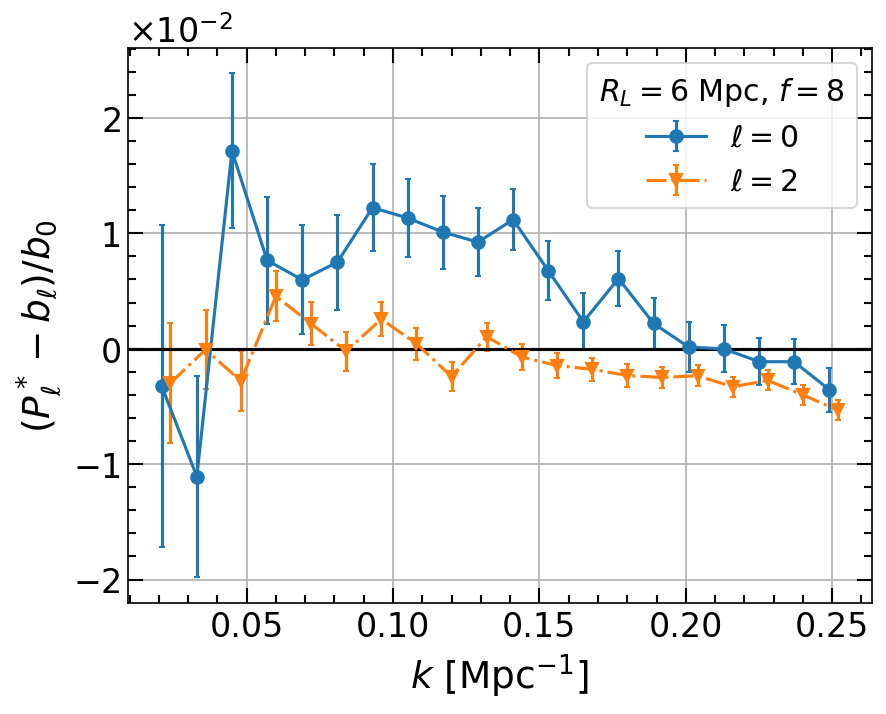}
    \hfill
    \includegraphics[width=0.482\linewidth]{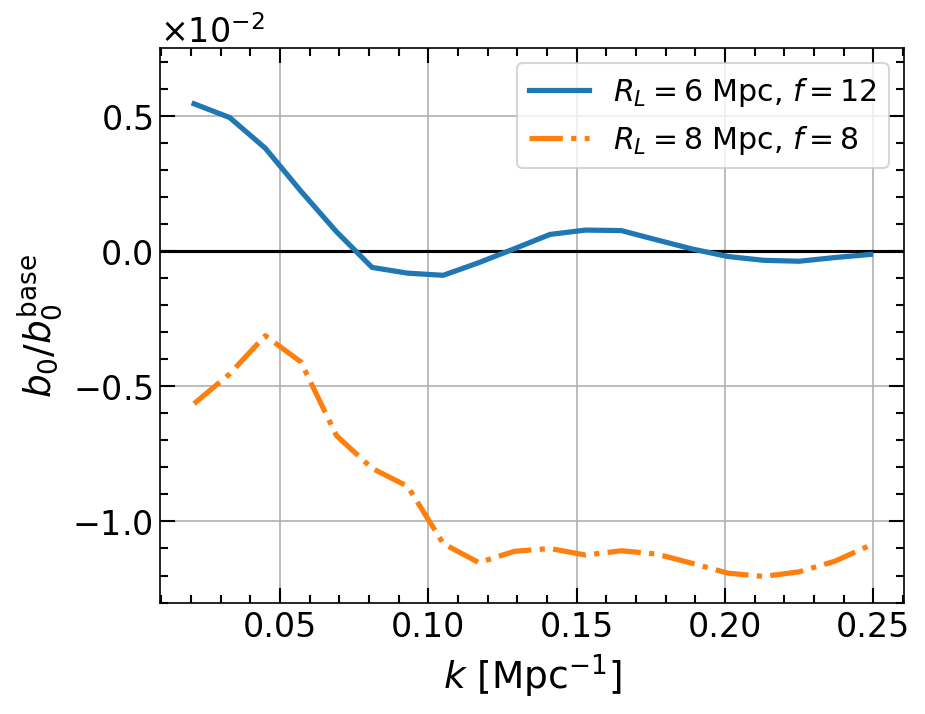}
    \caption{({\it Left}) Error in unnormalized multipole $P^*_\ell$ and total bias $b_\ell$ estimates (as defined in the text) relative to $b_0$ for the baseline setting $R_L=6~$Mpc and $f=8$. Note that these points are highly correlated according to the Fisher matrix. Nevertheless, the estimator seems to be biased at the percent level. ({\it Right}) Estimated total monopole bias $b_0$ for different large and small scale settings relative to the baseline monopole bias $b_0^\mathrm{base}$. Increasing the neighboring radius factor from $f=8$ to $f=12$ changes the total bias estimate at a sub-percent level on large scales. Increasing the smoothing scale $R_L$ to $8~$Mpc decreases $b_0$ on all scales by 1\% on average.}
    \label{fig:vary_rl_f_compare_bias}
\end{figure}

Across all tests, we find the estimator to be biased at the percent level, leading to high chi-squared values. This bias is illustrated for the baseline setting in the left panel of figure~\ref{fig:vary_rl_f_compare_bias} as the error in the unnormalized multipole $P^*_\ell(k) = \bm{d}^\mathrm{T} \mathbf{C}^{-1}\mathbf{C}_{,\ell k}\mathbf{C}^{-1} \bm{d}$ and total bias $b_\ell(k) = \mathrm{Tr}(\mathbf{C}^{-1}\mathbf{C}_{,\ell k})$ estimates relative to $b_0$ (these are the terms inside parentheses in eq.~\eqref{eq:p3d_estimator}).
According to the Fisher matrix, these points are highly correlated yet show a clear bias in the estimator. 

There are two possible sources for this bias. (1) A consistent hard boundary between large and small scales cannot be imposed in a \pcm\ code. Mathematically, all quasar pairs have to be included for the decomposition to be exact ($f\rightarrow\infty$). The missing quasar pairs due to a finite choice of $f$ could bias the estimator.  We investigate this below. (2) The density field has exactly the correct \pthreed\ on the mesh, regularly sampled points, which are interpolated to data locations. The unsmoothness of trilinear interpolation introduces an anisotropic suppression of power and ringing to the interpolated density field \cite{bernsteinResamplingImagesFourier2014}. These dominate at the Nyquist frequency $k_\mathrm{Nyq}\approx 3~$\impc; however, they could leave percent-level artifacts on the scales we consider here. The investigation of higher order interpolation kernels is left to future work.

In real-data applications, tuning the estimator’s parameters will be necessary. The error can be corrected using simulated data or mitigated by adjusting the estimator configuration.

\subsubsection{Varying large- and small-scale split factors}
We conduct two variations on the baseline setting to evaluate the impact of hyperparameters: the large-scale radius $R_L$ and the neighboring radius factor $f$. The impact of these variations on the total monopole bias $b_0$ is shown in the right panel of figure~\ref{fig:vary_rl_f_compare_bias}.

Keeping the large-scale radius fixed while increasing the neighboring radius factor to $f=12$ raises the average number of neighbors to 23 and changes $b_0$ at the sub-percent level on large scales. This change on large scales is expected since correlations from farther forests contribute to the bias calculation. However, the increased number of neighbors substantially impacts the computation time, increasing it by a factor of 3.5. Though the chi-squared value slightly decreases to 185 from 192 per 100 degrees of freedom, it remains high.

Increasing the large-scale radius to $R_L=8~$Mpc while keeping the neighboring radius factor fixed raises the average number of neighbors to 18 and decreases $b_0$ by approximately 1\% across all scales. The chi-squared value becomes even larger. This indicates that the PM component captures non-negligible small-scale information.

\subsubsection{Adding continuum errors and assessing its impact\label{subsubsec:continuum}}
We evaluate the efficacy of continuum marginalization using Gaussian simulations by emulating the impact of continuum fitting on the forest spectrum with a simple projection. Continuum marginalization has proven effective for the 1D power spectrum \cite{karacayliOptimal1DLy2020, karacayliOptimal1dDesiEdr2023}; however, the continuum fitting procedure itself introduces higher-order correlations, which we do not address here. Simulating realistic quasar continua and reproducing the eBOSS quasar continuum fitting pipeline are beyond the scope of this work.
\begin{figure}
    \centering
    \includegraphics[width=0.98\linewidth]{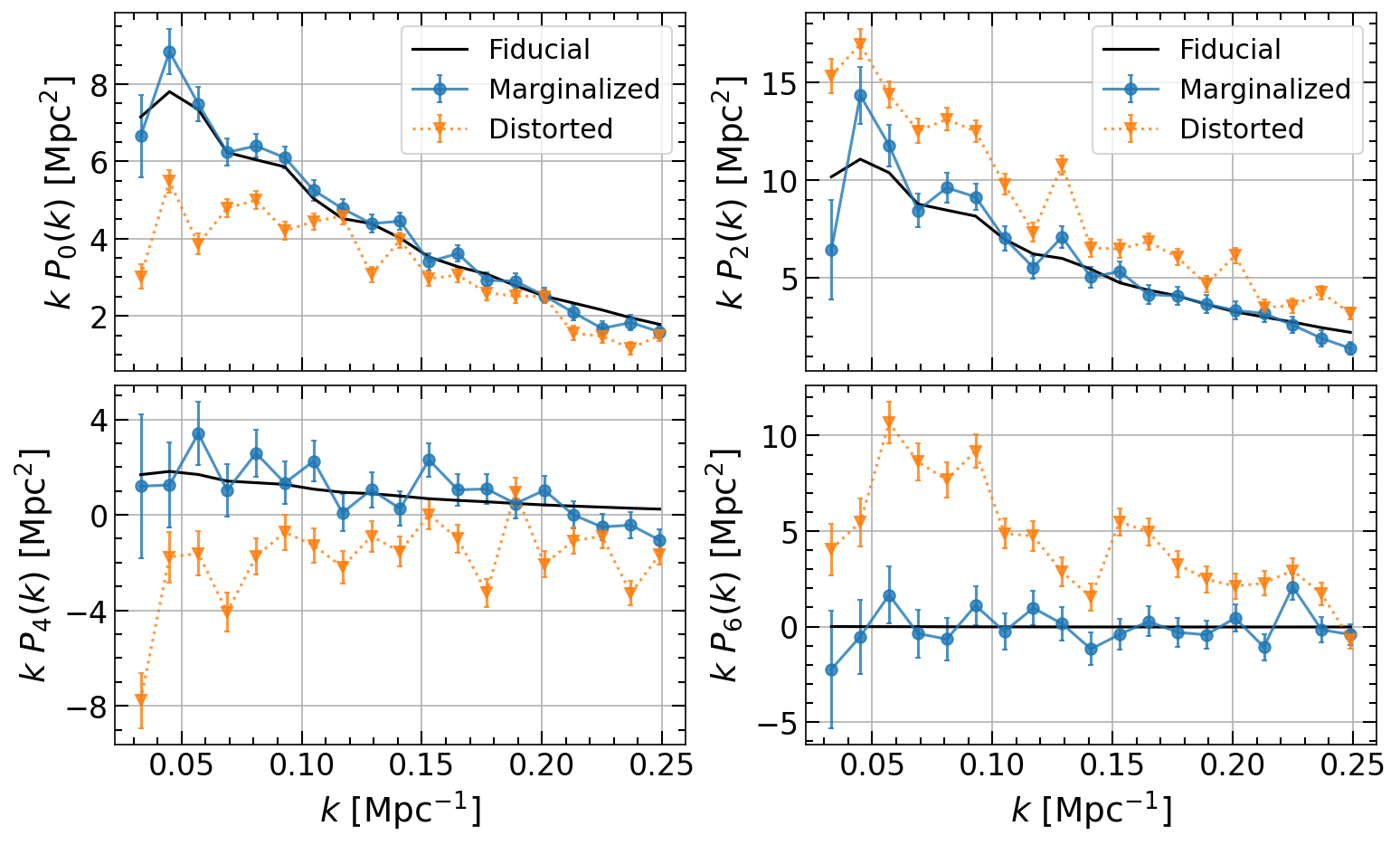}
    \caption{The distorted \pthreed\ ({\it orange dotted lines with triangles}) after projecting out the mean and the slope of $\delta_{F}$. The distortions suppress the monopole on large scales, boost the quadrupole on all scales, and extend power to higher-order multipoles ($\ell \geq 6$). The continuum marginalized \pthreed\ ({\it blue solid lines with circles}) agrees with the input fiducial model ({\it black solid lines}).}
    \label{fig:dist_p3d}
\end{figure}

The eBOSS continuum fitting is performed in the forest region following methodologies developed in refs.~\cite{bautistajuliane.MeasurementBaryonAcoustic2017, blomqvistBaryonAcousticOscillations2019}, which has also been applied to 1D studies~\cite{chabanierOnedimensionalPowerSpectrum2019, ravouxFFTP1dEDR2023, karacayliOptimal1dDesiEdr2023}. In this method, the definition of the quasar continuum includes the mean transmission $\overline{F}(z)$ of the IGM, and each quasar ``continuum" $\overline{F} C_q(\lambda_\mathrm{RF})$ is modeled by a first-order polynomial scaling of a global continuum: $\overline{F}C_q(\lambda_\mathrm{RF}) = \overline{C}(\lambda_\mathrm{RF}) \left( a_q + b_q \Lambda \right)$, where $\lambda_\mathrm{RF}$ is the wavelength in the quasar's rest frame and $\Lambda \sim \ln \lambda_\mathrm{RF}$. Since the fitting is done in the forest region, the resulting forest has, to a good approximation, a zero mean $\langle \delta_F \rangle = 0$ and a zero slope $\langle \Lambda \delta_F \rangle = 0$.

We project out these modes from $\delta_{F}$ in every forest to imitate continuum fitting errors. This projection is coupled to the underlying large-scale density mode and introduces distortions to 3D correlations. Figure~\ref{fig:dist_p3d} illustrates the distorted multipoles in orange dotted lines. This parallels the findings of ref.~\cite{belsunce3dLymanAlphaPowerSpectru2024}: The monopole is suppressed on large scales, and the quadrapole is boosted on all scales. Additionally, distortions spread spurious signals to higher-order multipoles, which is undesired as it makes theoretical modeling difficult. As an example, we plot $P_{\ell = 6}(k)$ in the bottom right panel of figure~\ref{fig:dist_p3d}.

The continuum marginalization, described in section~\ref{subsec:continuum}, removes these distortions from the measurement, shown in blue solid lines in figure~\ref{fig:dist_p3d}. The first three main multipoles now agree with the input model, and the large distorted signal of $\ell = 6$ is completely removed. However, as established previously with MC simulations, this marginalization diminishes the precision of measured \pthreed\ and correlates higher-order multipoles.

\section{Discussion\label{sec:discussion}}

We have developed an optimal estimator for the 3D power spectrum (\pthreed) of the \lya\ forest, designed to overcome the unique challenges posed by its highly anisotropic, uneven sampling and continuum fitting errors that distort the two-point functions. Our approach builds on optimal estimator techniques of cosmic microwave background (CMB) data. It is statistically optimal (for Gaussian fields; in the presence of non-Gaussianities, it remains unbiased but is not, strictly speaking, optimal), deconvolves the survey window function, and marginalizes contaminated modes. The current implementation works in the flat-sky approximation.

To make the optimal estimator computationally feasible for large datasets, we implemented a particle-particle/particle-mesh (\pcm) decomposition of the covariance matrix. This allows us to efficiently handle large-scale correlations using fast Fourier-space convolution while computing small-scale correlations via direct summation. We validated our implementation using Monte Carlo mocks and Gaussian simulations, and demonstrated that our estimator is computationally viable, albeit slightly sensitive to hyperparameter choices such as the large-scale split scale $R_L$ and neighboring radius factor $f$. We found the estimator to be biased at the percent level, which must be corrected using simulated data or mitigated by tuning the estimator configuration in real data applications. Higher order interpolation kernels could reduce this bias \cite{bernsteinResamplingImagesFourier2014}, which we leave to future work.

The current estimator successfully handles several complexities of the \lya\ forest. We addressed redshift evolution by computing the signal matrix at a pivot redshift and applying a separable redshift growth function. By propagating this to derivative matrices, we were able to estimate \pthreed\ at the pivot redshift, removing the mixing of different time periods. Our redshift evolution treatment handles only the amplitude, which is valid for linear scales, but invalid for non-linear scales. However, these non-linear scales are better probed by densely sampled surveys and by using a higher-resolution mesh with a shorter radial distance, in which case the error of this simplified treatment will be alleviated. One could overcome this shortcoming by using multiple grids at different redshifts and interpolating them. Additionally, our single-grid approach assumes metals and DLAs grow similarly to the \lya\ forest, which may require refinement in future studies, such as employing separate grids for these systems.

We also implemented marginalization over quasar continuum-fitting degeneracies in the weighting of the estimator rather than on the data vector itself. This means that we do not have to include the continuum-fitting distortion of the theory; instead, the loss of information manifests as increased errors on large scales and increased correlations between multipoles. Specifically, the positive correlations between $\delta \ell = 2$ modes change to anti-correlations, and correlations extend to $\delta \ell \geq 8$ after mode marginalization.

Our estimator is most effective at recovering large-scale information when applied to sparsely sampled large-scale structure surveys. Accurate access to these large-scale modes enables robust interpretations of Alcock–Paczyński distortions, ionizing background fluctuations, and signatures of primordial non-Gaussianity.  In addition, BAO can be measured in power spectrum space using a peak/smooth template decomposition, complementing \lya\ forest correlation function analyses \cite{desiKp6BaoLya2024, desiY3LyaBAO2025}. The BAO feature is delocalized in Fourier space and appears as oscillations that must be densely sampled in $k$ bins to avoid smearing. A key advantage of an optimal power spectrum measurement is its robustness against survey window complications, which pose an important challenge in galaxy power spectrum BAO analyses \cite{beutlerGalaxyBaoFourierSdss2017}.


Our next step toward application to real data will be to implement spherical geometry. Initial analyses of eBOSS data and mocks that are generated on a spherical geometry revealed that the estimator is strongly biased by geometric distortions. These effects, coupled with continuum error modes, contaminate the measurement beyond recovery. Though tuning the estimator settings, such as estimating \pthreed\ in smaller sky patches and narrower redshift bins, can mitigate this effect, we deem reformulating the estimator in spherical geometry as the most robust solution.

The generalization to a spherical sky will require significant additions to the PM part of the code (and more modest changes to the P$^2$ algorithm). Spherical geometry requires the replacement of the 3D FFTs with spherical harmonics for the angular directions and spherical Bessel functions for the radial directions. The 3D Fourier transform has cost ${\cal O}(N\log N)$, where $N = N_x N_y N_z$ is the number of mesh points. The spherical convolution on a mesh requires several steps. First is an FFT in the longitude direction to replace right ascension $\phi$ with angular momentum quantum number $m$, with cost ${\cal O}(N_r N_\theta N_\phi \log N_\phi)$, where $N_r$, $N_\theta$, and $N_\phi$ are the number of mesh points in the radial, declination, and right ascension directions, respectively. Second is the replacement of declination $\frac\pi2-\theta$ with angular momentum quantum number $\ell$, with cost ${\cal O}(N_r N_\theta^2 N_\phi)$ if done using standard recursion relations (as done in, e.g., the spherical PM implementation in SDSS analyses \cite{hirataCrossCorrelationCmbLss2004}). Then comes the transformation of radial distance $r$ to wavenumber $k$: this has cost ${\cal O}(N_r^2 N_\theta N_\phi)$ if done by direct multiplication by the spherical Bessel functions $j_\ell(kr)$. In principle, one could use a logarithmic mesh in $r$ and do an FFT-based convolution in the radial direction \cite{hamiltonUncorrelatedModesNonlinear2000}, but given the narrow range in $\log r$ probed by the \lya\ forest, it is not clear whether we want to pursue this approach. The inclusion of redshift-space distortions also requires multiplication by matrices involving the derivatives of $j_\ell(kr)$ \cite{fisherSphericalHarmonicRsd1994} or equivalently the adjacent $j_{\ell\pm 2}(kr)$ \cite{padmanabhanClusteringLrgSdss2007}, which fortunately can be computed from the same recursion relations as the $j_\ell$'s. If $N_r$, $N_\theta$, and $N_\phi$ are of the same order of magnitude, then, we expect the spherical implementation to have cost ${\cal O}(N^{4/3})$, which is slower than the FFT but by the modest and (hopefully) manageable factor $\sim N^{1/3}/\log N$.

The estimator's sensitivity to the large-scale split scale $R_L$ and the search radius for neighbors $f R_L$ necessitates careful calibration through simulations tailored to the specific dataset. Specifically, the parameter $f$ describes the overlap region between the large- and small-scale treatments. There must be some transition region in any \pcm\ code since the large scales are handled in Fourier space, the small scales in configuration space, and one cannot consistently impose a hard boundary in both spaces. However, increasing $f$ significantly raises computational costs, making it impractical beyond certain limits. On the other hand, performing the PP component on GPUs could substantially improve the efficiency of the small-scale matrix-vector multiplications and allow more distant neighbors to be included. Exploring GPU implementations to optimize performance is a promising avenue for future work.

One possible drawback of our optimal estimator is that it auto-correlates spectra by construction in contrast to the estimators of refs.~\cite{fontriberaEstimate3DPowerLya2018, belsunce3dLymanAlphaPowerSpectru2024}. The auto-correlations lead to added spectral noise power and are susceptible to pipeline calibration errors and sky modeling errors \cite{guyCharacterizationOfContaminantsDesi2024}. Furthermore, residuals in the sky background subtraction model correlate pixels between neighboring quasars observed in the same spectrograph \cite{guyCharacterizationOfContaminantsDesi2024}. The cosmology inference must include templates to marginalize errors in noise calibration and sky modeling.
We show in appendix~\ref{app:quasi_cross} how the estimator can be altered to calculate only cross-forest contributions, and discard some combinations of forest pairs, albeit sub-optimally.

This work represents a significant step forward in the precise measurement of \pthreed, paving the way for more accurate cosmological constraints from upcoming large-scale structure surveys like DESI and Spec-S5 \cite{specS5}.

\appendix
\section{Matrix square root \texorpdfstring{$\mathbf{Q} = \mathbf{A}^{1/2}$}{Q=Sqrt(A)}\label{app:matrix_sqrt}}
Ref.~\cite{highamStableIterationsMatrixSquareRoot1997} is a great resource on matrix square root algorithms, and we have heavily relied on it below. 

\subsection{Newton iteration}
The ordinary Newton iteration is as follows: $\mathbf{Q}_{i+1} = (\mathbf{Q}_i + \mathbf{A} \mathbf{Q}_i^{-1}) / 2$.
Starting the iterations with the block-diagonal matrix, $\mathbf{Q}_0 = \mathbf{A}_\mathrm{BD}^{1/2}$, we arrive at the following expression for the second iteration:
\begin{equation}
    \mathbf{Q}_2 = \frac{1}{4} \mathbf{Q}_0 + \mathbf{H} \mathbf{Q}_0 \left(\frac{1}{4}\mathbf{I} + (\mathbf{I} + \mathbf{H})^{-1} \right),
\end{equation}
defining $\mathbf{H} \equiv \mathbf{A} \mathbf{A}_\mathrm{BD}^{-1}$. We find that the matrix $\mathbf{H}$ has a small condition number. The inverse matrix $(\mathbf{I} + \mathbf{H})^{-1}$ can be solved with the conjugate gradient method, which has a faster convergence property than the full covariance matrix, thanks to the additional identity matrix. Note that $\mathbf{H}$ is not symmetric. Nevertheless, the conjugate gradient method rapidly converges. Unfortunately, subsequent iterations require nested matrix inverses and are prohibitively expensive. Furthermore, this formulation of the Newton iteration is actually unstable \cite{ highamStableIterationsMatrixSquareRoot1997}. Both of these factors render this method impractical and unworthy of further pursuit.

\subsection{Newton-Schulz iteration}
The Newton-Schulz (NS) iteration is a commonly used alternative that is more stable but requires two coupled matrices \cite{schulzIterativeMatrix1933, highamStableIterationsMatrixSquareRoot1997, ChallacombeNbodySolverSqrt2015, songFastDifferentiableMatrixSqrt2022}. Starting from $\mathbf{Q}_0 = \mathbf{A}$ and $\mathbf{Z}_0 = \mathbf{I}$, the coupled iteration is as follows:
\begin{equation}
    \mathbf{Q}_{i + 1} = \frac{1}{2} \mathbf{Q}_i (3\mathbf{I} - \mathbf{Z}_i \mathbf{Q}_i), \qquad \mathbf{Z}_{i + 1} = \frac{1}{2} (3\mathbf{I} - \mathbf{Z}_i \mathbf{Q}_i) \mathbf{Z}_i.
\end{equation}
Provided proper normalization (e.g., $\mathbf{A}\rightarrow \mathbf{A} / \lambda_\mathrm{max}$, $\mathbf{A}\rightarrow \mathbf{A} / \mathrm{max}(A_{ii})$ or $\mathbf{A}\rightarrow \mathbf{A} / \|\mathbf{A}\|_F$), the NS iteration has quadratic convergence of $\mathbf{Q}_i$ to $\mathbf{A}^{1/2}$ and $\mathbf{Z}_i$ to $\mathbf{A}^{-1/2}$. Unlike the ordinary Newton iteration, the initial values cannot be altered in the NS method. Additionally, the NS method does not involve inverting matrices. But this advantage disappears as we cannot store either $\mathbf{Q}_i$ or $\mathbf{Z}_i$ matrices. We implemented the NS iteration as recursively nested multiplications, storing the intermediate resulting vectors. For $N_i$ iterations, this requires $2\times N_i$ additional temporary vectors. For low values of $N_i$, the additional memory requirement remains manageable. However, the number of matrix multiplications increases rapidly with each iteration, reaching 365 multiplications for $N_i=6$.

\subsection{Padé iteration}
The Padé iteration is also derived from the matrix sign function like NS iteration \cite{KennyLaubRationalIterativeMatrix1991, pandeyParalelAlgorithmMatrixSign1990, KennyLaubHyperbolicPade1994, highamStableIterationsMatrixSquareRoot1997}. For a Padé approximation order $p$, the coupled iteration for the matrix square root is the following:
\begin{equation}
    \mathbf{Q}_{i + 1} = \frac{1}{p} \mathbf{Q}_i \sum_{i=1}^p \frac{1}{\xi_i} ( \mathbf{Z}_i  \mathbf{Q}_i  + \alpha_i^2\mathbf{I})^{-1}, \qquad \mathbf{Z}_{i + 1} = \frac{1}{p} \mathbf{Z}_i \sum_{i=1}^p \frac{1}{\xi_i} ( \mathbf{Q}_i \mathbf{Z}_i  + \alpha_i^2\mathbf{I})^{-1},
\end{equation}
where $\xi_i=\left(\cos((2i-1)\pi/2p) + 1\right) / 2$ and $\alpha_i^2=1/\xi_i - 1$. The iterations start from $\mathbf{Q}_0 = \mathbf{A}$ and $\mathbf{Z}_0 = \mathbf{I}$ and cannot be altered like the NS method. The nested matrix inverses are still prohibitively expensive; however, unlike ordinary Newton iteration, the accuracy of the Padé approximation can be increased arbitrarily by $p$.  Additionally, increasing $p$ does not require additional memory. While computing $p$ matrix inverses may initially appear excessively costly, their convergence is accelerated by the addition of $\alpha_i^2$ to the diagonal. We find that for $p=4$, all four matrix inverses converge in fewer than 40 iterations in total for $\epsilon=10^{-5}$. Ref.~\cite{highamStableIterationsMatrixSquareRoot1997} demonstrates that $p=4$ converges within a few iterations across all scenarios, further supporting our single-iteration formulation.

\section{Quasi-optimal cross-forest estimator\label{app:quasi_cross}}
QMLE formalism can be extended to construct an unbiased cross-forest estimator that partially preserves inverse covariance weighting. However, this estimator does not inherit the theoretical minimum variance property of the full optimal estimator as discussed also in the context of CMB anisotropies \cite{planckIntermediateResults2016}.

First, to prevent different forests from mixing, the inverse covariance weighting must be isolated to individual forests through block-diagonal approximation. This is a common approximation in previous \lya\ analyses \cite{slosarMeasurementBaryonAcoustic2013, metcalfLyaWeakLensing2020} but leads to sub-optimality. Second, only different sightlines must be correlated in the derivative matrix operation. Putting these together, we can start formulating the estimator as follows:
\begin{equation}
    y^*_k = \frac12 \bm{d}^\mathrm{T} \mathbf{C}_\mathrm{BD}^{-1} \mathbf{C}^X_{,k} \mathbf{C}_\mathrm{BD}^{-1} \bm{d} - b^X_k,
\end{equation}
where $\mathbf{C}^X_{,k'}$ is non-zero only for off-block-diagonal terms and symmetric. This is unbiased if $\langle y^*_k \rangle = 0$, which means 
\begin{equation}
    b^X_{k}=\frac12 \Tr\left( \mathbf{C}_\mathrm{BD}^{-1} \mathbf{C}^X_{,k} \mathbf{C}_\mathrm{BD}^{-1} \mathbf{C} \right).
\end{equation}
Then, we can write the quasi-optimal cross-forest estimator analogous to eq.~\eqref{eq:p3d_estimator}:
\begin{equation}
    \hat P^\mathrm{3D}_{k} = P^\mathrm{3D, fid}_{k} + \sum_{k'} \frac{1}{2} F^{-1}_{kk'}\left[\bm{d}^\mathrm{T} \mathbf{C}_\mathrm{BD}^{-1} \mathbf{C}^X_{,k'} \mathbf{C}_\mathrm{BD}^{-1} \bm{d} -  b^X_{k'} \right],
\end{equation}
and
\begin{equation}
    F_{kk'} = \frac12 \Tr(\mathbf{C}_\mathrm{BD}^{-1} \mathbf{C}^X_{,k} \mathbf{C}_\mathrm{BD}^{-1} \mathbf{C}^X_{,k'}).
\end{equation}

However, the covariance matrix of the estimated $\hat P^\mathrm{3D}_{k}$ is not simply the inverse Fisher matrix. Denoting the covariance of $y^*_k$ as $V^{(y)}_{kk'} \equiv \langle y^*_k y^*_{k'} \rangle$, the covariance of estimated $\hat P^\mathrm{3D}_{k}$ is given by $\mathbf{V}^{(P)} = \mathbf{F}^{-1} \mathbf{V}^{(y)} \mathbf{F}^{-1}$. The stochastic trace estimation of $b_k^X$ will give $\mathbf{V}^{(y)}$ for free as before. The random vectors with correlated sightlines can be drawn with our prescription in section~\ref{subsec:estimate_bias_fisher}. In case direct estimation is desired, $\mathbf{V}^{(y)}$ is analytically given by
\begin{equation}
    V^{(y)}_{kk'} = 2 \Tr \left(\mathbf{C} \mathbf{E}_k \mathbf{C} \mathbf{E}_{k'} \right)
\end{equation}
under the Gaussianity assumption, where we defined $\mathbf{E}_k \equiv (1/2) \mathbf{C}_\mathrm{BD}^{-1} \mathbf{C}^X_{,k} \mathbf{C}_\mathrm{BD}^{-1}$ for clarity.



This estimator does not require a conjugate gradient method solution, which is a significant speed improvement over the optimal estimator. However, the derivative matrix operation can be unfeasible if every forest pair needs to be multiplied. This operation can be performed efficiently by utilizing $\mathbf{C}^X_{,k} = \mathbf{C}^\mathrm{Total}_{,k} - \mathbf{C}^\mathrm{BD}_{,k}$, such that the operation $\mathbf{C}^\mathrm{Total}_{,k}$ can be performed using the PM formalism from which the block-diagonal contributions can be subtracted. This formalism can be extended to explicitly remove some combinations of forest pairs, such as neighboring quasars of the same spectrograph, from $\mathbf{C}^X_{,k}$. 

\section{Hermite interpolation details\label{app:hermite}}
The most publicly accessible reference for cubic Hermite interpolation can be found online in Wikipedia,  the free encyclopedia \cite{wikipediaCubicHermiteSpline2025}. Here we lay out our formulation and cite the historical references that contribute to the efficient calculation of this interpolation scheme. A central reference we recommend is a book by Farin \cite{farin2001curves}. We will assume the input points $x_k$ are regularly spaced with $\Delta x$ spacing. The interpolating cubic Hermite polynomial between points $x_k$ and $x_{k+1}$ is given by
\begin{equation}
    y(x) = y_k B^3_0(t) + (y_k + y'_k\Delta x/3) B_1^3(t) + (y_{k+1} - y'_{k+1}\Delta x/3)B_2^3(t) + y_{k+1} B_3^3(t),
\end{equation}
where $t\equiv(x-x_k)/\Delta x$ and $B_i^n(t) = \left(
\begin{array}{c}
n\\
i\\
\end{array}
\right) t^i(1-t)^{n-i}$ are the Bernstein polynomials \cite{bernstein1912}. As we will soon highlight, these polynomials are not explicitly evaluated. We use the finite-difference method for the derivatives, which becomes a Catmull-Rom spline \cite{CATMULL1974317} for regular grids: $y'_k = (y_{k+1} - y_{k-1})/2\Delta x$. Putting this into the equation above, we see that the four nearest points of the input are used in evaluation.

This polynomial can be efficiently evaluated using de Casteljau's recursive algorithm \cite{deCasteljau1959outillages} as follows. First, we write $y(x) = \sum_{i=0}^3 \beta_i B^3_i(t)$, where $\beta_i$ are the coeffients of the interpolating polynomial. The algorithm recursively updates these coefficients. For iteration $r$, the new coeffients are $\beta^r_i = \beta^{r-1}_i (1-t) + \beta_{i+1}^{r-1}t$ where $i=0,\dots, 3-r$ and $r=1,2,3$. The interpolation result is $y(x) = \beta_0^3$. This efficient algorithm performs six operations, each with two multiplications and one addition, and does not evaluate Bernstein polynomials.

\acknowledgments

We thank Nikhil Padmanabhan for the fruitful discussion.

N.G.K. acknowledges support from the United States Department of Energy, Office of High Energy Physics under Award Number DE-SC0011726.

During the preparation of this work, C.H. was supported by the David \& Lucile Packard Foundation; and by the National Aeronautics and Space Administration (grant 22-
ROMAN11-0011).


\paragraph{Software.} Our estimator is written in \texttt{c++}.\footnote{\url{https://github.com/p-slash/lyspeq}}
It depends on \texttt{CBLAS} and \texttt{LAPACKE} routines for matrix/vector operations, \texttt{GSL}\footnote{\url{https://www.gnu.org/software/gsl}} for certain scientific calculations \citep{GSL}, \texttt{FFTW3}\footnote{\url{https://fftw.org}} for fast Fourier transforms \citep{FFTW05}; and uses the Message Passing Interface (MPI) standard\footnote{\url{https://www.mpich.org}} to parallelize tasks.
We use the following commonly-used software in \texttt{python} analysis: \texttt{astropy}\footnote{\url{https://www.astropy.org}}
a community-developed core \texttt{python} package for Astronomy \citep{astropy:2013, astropy:2018, astropy:2022},
\texttt{numpy}\footnote{\url{https://numpy.org}}
an open source project aiming to enable numerical computing with \texttt{python} \citep{numpy},
\texttt{scipy}\footnote{\url{https://scipy.org}} an open-source project with algorithms for scientific computing.
Finally, we make plots using
\texttt{matplotlib}\footnote{\url{https://matplotlib.org}}
a comprehensive library for creating static, animated, and interactive visualizations in \texttt{python}
\citep{matplotlib}.

\bibliographystyle{JHEP}
\bibliography{biblio.bib}

\end{document}